\documentclass{aa}  
\usepackage{graphicx}
\usepackage{txfonts}

\usepackage{amsmath}

\begin{document}

   \title{Synergies between the COMAP CO Line Intensity Mapping mission and a Ly$\alpha$ galaxy survey: How to probe the early universe with voxel based analysis of observational data}

   \titlerunning{Synergies between COMAP and a Ly$\alpha$ galaxy survey}

   \author{Marta B. Silva\inst{1}  
        \thanks{Corresponding author: Marta B. Silva, e-mail: martabrunosilva@gmail.com}
          \and Bernhard Baumschlager\inst{1} 
          \and Kieran A.~Cleary \inst{2} 
          \and Patrick C.~Breysse \inst{3} %
          \and Dongwoo T.~Chung \inst{4,5} %
          \and H\aa vard T. Ihle \inst{6} %
          \and Hamsa Padmanabhan \inst{7} 
          \and Laura C.~Keating \inst{8} 
          \and Junhan Kim \inst{2} 
          \and Liju Philip \inst{9} 
   }
   \authorrunning{COMAP Collaboration}

   \institute{Institute of Theoretical Astrophysics, University of Oslo, P.O. Box 1029 Blindern, N-0315 Oslo, Norway
    \and California Institute of Technology, 1200 E. California Blvd., Pasadena, CA 91125, USA
    \and Center for Cosmology and Particle Physics, Department of Physics, New York University, 726 Broadway, New York, NY, 10003, USA
    \and Canadian Institute for Theoretical Astrophysics, University of Toronto, 60 St. George Street, Toronto, ON M5S 3H8, Canada
    \and Dunlap Institute for Astronomy and Astrophysics, University of Toronto, 50 St. George Street, Toronto, ON M5S 3H4, Canada
    \and D\'epartement de Physique Théorique, Universite de Genève, 24 Quai Ernest-Ansermet, CH-1211 Genève 4, Switzerland
    \and Leibniz-Institut f{\"u}r Astrophysik Potsdam (AIP), An der Sternwarte 16, D-14482 Potsdam, Germany
    \and Jet Propulsion Laboratory, California Institute of Technology, 4800 Oak Grove Drive, Pasadena, CA 91109, USA\\
    }

   \date{Received / Accepted }

 
  \abstract
  {Line Intensity Mapping (LIM) offers a novel avenue to observe and characterize our universe. This technique can provide a global context to observations by traditional galaxy surveys and can revolutionize our knowledge of cosmology. LIM data using spectral lines of metals are becoming available, such as those obtained by the CO Mapping Array Project (COMAP), which started observing in 2019 with a Pathfinder experiment. COMAP data can be used to probe the molecular gas content of the universe from the last stages of the Epoch of Reionization (EoR) ($z<8.0$) to the epoch of stellar build up ($z\sim 2.5$).}
  {In this work, as an alternative to auto- or cross-power spectrum analyses, we examine the prospects for deriving voxel-level statistical constraints on high-redshift galaxies from COMAP data by considering the additional information available from a galaxy survey.
  For the galaxy survey, we consider observations using the Visible Integral-Field Replicable Unit Spectrograph (VIRUS) on the Hobby-Eberly Telescope (HET). This instrument will survey part the COMAP volume as part of the HET Dark Energy Experiment (HETDEX) survey of Lyman alpha emitting (LAE) galaxies and we also consider follow-up observations with a higher fill-factor and greater overlap.}
  {We post-process the IllustrisTNG300 galaxy-formation simulation with a set of prescriptions to consistently determine CO and Ly$\alpha$ line luminosities. The different line prescriptions span the uncertainty in the CO line luminosity according to current observations by VLA and ALMA high-$z$ CO surveys and set the Ly$\alpha$ emission to be compatible with observational LAE luminosity functions. We produce consistent mock observations for the two surveys over a $\sim300^3\, {\rm Mpc^3}$ volume. These are  then used to formulate and test methodologies for data analysis and to predict COMAP constraints on CO emission. We use combinations of masking, stacking, voxel intensity distribution (VID), and other statistics.}
   {We find that in combination with HETDEX, a voxel-level analysis of the COMAP Pathfinder survey can detect and characterize the CO signal from $z\sim3$ and improve current constraints on the $z\sim6$ signal, identify individual voxels with bright CO(1-0) emission at $z\sim3$ and probe the redshift evolution of the CO emission and potentially the CO line width. Moreover, with the two data-sets we can identify relationships between the two lines driven by dust.}
   {This study illustrates the potential of synergies between LIM and galaxy surveys both to improve the significance of a detection and to aid the interpretation of noisy LIM data.}

   \maketitle 
%

\section{Introduction} \label{section1}

Galaxies are biased tracers of the large scale structure of the Universe and their properties and evolution are largely influenced by their global environment. A series of cosmology missions such as SKA \citep{2020PASA...37....7S}, Euclid \citep{2018LRR....21....2A} and HETDEX (\cite{2008ASPC..399..115H}; Hill, G.J. et al 2021, AJ submitted; Gebhardt, K., et al 2021, ApJ submitted) propose to map large-scale structure over a large fraction of cosmic time using galaxies as tracers of matter density. Each of these missions will probe bright galaxies using specific spectral lines or limited frequency bands and so they will only detect a fraction of the galaxies over the observed volume. The connection between matter density and observed galaxy number density and spatial distribution depends on how representative the galaxy survey selection is of the underlying matter. Consequently, there is a significant level of uncertainty in the bias between observed signals and the matter power spectra.

Our understanding of the global properties of galaxies is based on biased observations of a limited number of sources. 
Current and next-generation galaxy surveys are taking steps to overcome these limitations by pushing observations towards earlier times, larger volumes and by performing blind surveys. However, due to sensitivity limits and high observing costs, galaxy surveys cannot reach all of these goals at once.

Line intensity mapping is a novel observational technique which was first developed to detect radio HI 21cm line emission from gas in the inter-galactic medium (IGM) during the Epoch of reionization and from galaxies at lower redshifts (usually below $ z\sim 3$) \citep{1997ApJ...475..429M}.
Later this technique was extended to probe galaxies using bright hydrogen UV lines such as Lyman alpha and H-alpha and metal lines such as CII and CO rotation lines. LIM of galaxies (usually referring to emission lines other than HI 21cm) can complement traditional galaxy surveys by performing unbiased, wide and deep surveys of emission from unresolved sources over a large fraction of the age of the Universe. This technique provides a rich data-set that can revolutionize our knowledge of cosmology and global galaxy properties. The first observational data-sets and results from LIM surveys targeting metal lines are only now becoming available. These pathfinder surveys provide the first non HI LIM data-sets and require dedicated processing pipelines and analysis tools to be developed and optimized. 

As a first step, LIM analysis takes advantage of tools developed to process observations of the cosmic microwave background or from galaxy surveys. However the former signal has a solid theory framework to guide the analysis pipeline and the latter provides information about individual systems which can to first order be analyzed on its own. Line intensity mapping observations have none of these characteristics; hence these missions' full potential can only be reached with the development of its own analysis tools. 

A key goal of these first LIM surveys is to detect the three-dimensional auto-power spectrum of the line emission from galaxies and use this measurement to constrain the galaxies' global properties. However, the scientific potential of LIM data is greatly enhanced through the combination of multiple data-sets, each probing different aspects of the galaxies. 
Such synergies can be found by performing cross-correlations between LIM data-sets to achieve or enhance detection, or by combining the data-sets to improve redshift precision, identify  line interlopers, etc. The large variety of possible data-sets to combine means that the specific analysis techniques required will need to be developed on a case-by-case-basis.   
In this work we explore a set of  tools for the analysis of data from the CO Mapping Array Project (COMAP) making use of information from a catalog of LAEs from the HETDEX galaxy survey. We will focus on a voxel-level analysis; a discussion of the potential for cross-power spectrum analysis between COMAP and HETDEX is discussed elsewhere \cite{2019ApJ...872..186C}. 

Observations with the COMAP Pathfinder started in 2019 (Cleary et al. in prep.), targeting ${\rm CO}(1-0)$ line emission at the epoch of stellar buildup ($z\sim 3$) and ${\rm CO}(2-1)$ from the late stages of the Epoch of Reionization $z \sim 6$.
The first CO rotation line is a good tracer of molecular gas in metal rich galaxies. Hence, COMAP lower-redshift data can be used to constrain the total molecular gas content at $z\sim3$ and to determine whether it peaks at the same time as the cosmic star formation rate density. The higher redshift ${\rm CO}(2-1)$ signal is more uncertain, given that we have a very limited knowledge of the properties of the galaxies sourcing this emission. This signal depends on molecular gas masses, dust content and other ISM gas properties. Simulating ${\rm CO}(2-1)$ emission at $z\sim 6$ is challenging due to the need to resolve molecular gas clouds and the required assumptions about several poorly constrained parameters, hence COMAP observational limits on this signal will have strong consequences for our knowledge of the global properties of this first generation of galaxies. 

The significance to which COMAP current and future surveys are expected to constrain the EoR CO signal are discussed in detail in a dedicated COMAP study (Breysse et al in prep.).   

Galaxy surveys made using ALMA and the VLA have put the first limits on the bright end of the CO luminosity function at $z>2$ \citep{2019Decarli,2019Riechers}. 
However, galaxy surveys are sensitivity-limited and so will miss the signal from faint sources. Given the overall high number-density of low-CO sources, if their brightness is high enough they will make a significant contribution to the overall CO intensity. 

COMAP measurements will constrain the total CO luminosity density and place the first limits on the low end of the CO luminosity function. Hence, COMAP surveys are complementary to CO galaxy surveys made using ALMA, VLA and the future ngVLA. COMAP surveys can cover large volumes with high redshift precision and so they are a potential cosmology probe. In addition, COMAP data can be combined and cross-correlated with external galaxy surveys, LIM surveys or cosmic microwave background (CMB) surveys.

COMAP surveys were chosen so as to overlap with the field covered by the HETDEX survey at $z\sim3$. HETDEX will perform blind spectroscopic survey to map Lyman alpha emitters (LAEs) over cosmological volumes.

The improved prospects for a signal detection through the cross-correlation of 3D maps from the two missions has been explored by \cite{2019ApJ...872..186C}. This study uses a prescription for CO line emission and LAE emission as a function of dark matter halo mass. 
Here we revisit the \cite{2019ApJ...872..186C} study using a set of updated line luminosity models that make it possible to predict line luminosity as a function of individual galaxy properties, consistent with the latest observational limits on the line luminosity functions. 
We produced mock COMAP observations by using the CO simulations from Baumschlager et al. (in prep.). The Baumschlager et al. simulations were obtained by post processing 3D maps from the Illustris TNG simulation \citep{2019ComAC...6....2N, 2018MNRAS.475..624N, 2018MNRAS.475..676S, 2018MNRAS.477.1206N, 2018MNRAS.475..648P, 2018MNRAS.480.5113M} using a series of semi-analytic prescriptions to derive first the galaxies' molecular gas content and later ${\rm CO}(1-0)$ and ${\rm CO}(2-1)$ luminosities. We adopt a set of three CO line models that cover the most likely range of the signals as predicted by both the simulations and by observations by the VLA COLDz survey \citep{2019ApJ...872....7R}.
Moreover, we use the same base simulations to consistently model the Lyman alpha emission in these galaxies and produce mock LAE observations for the HETDEX mission.
The LAE luminosity function at $z=3$ is fairly  well constrained by galaxy surveys such as MUSE \citep{2019A&A...624A.141U} and so we adjust the escape fraction of Lyman alpha photons in our model to fit current observations. The methodology to produce the CO and LAE mock maps is further described in Section \ref{sec:3}.

The main goal of this study is to explore the science constraints that can be obtained by voxel-level analyses of COMAP line intensity maps incorporating information about HETDEX galaxies. With that aim, we test a series of six analysis strategies to combine these data-sets that are robust to the uncertainty on the signal modelling and can be directly applied to noisy, observational CO maps. The explored analysis recipes are based on real-space voxel intensity data and were designed or adapted to fit the characteristics of the COMAP and HETDEX data. Most of these recipes can be adjusted and applied to other combinations of line intensity mapping and galaxy survey data.
The connection between our goals and the analysis techniques outlined in this paper is clarified in Section~\ref{sec:Analise_outcomes}.

\section{Analysis methods and science outcomes}
\label{sec:Analise_outcomes}

Tables \ref{tab2} and \ref{tab3} connect the science goals of the COMAP and HETDEX collaboration with the overall analysis techniques described in Section~\ref{sec:Analysis}. Table~\ref{tab2} contains the most important and easily achievable goals for this synergies study while Table~\ref{tab3} contains possible goals which are dependent on the actual spatial connection between the CO emission and the LAEs or on the signals' amplitude.

\begin{table*}
\centering
\caption{Science program and analysis methods: Main Goals}
\begin{tabular}{l | c | c | c} 
\hline\hline
    \textbf{Overall Science } &  \multicolumn{3}{c}{\textbf{Galaxy evolution from the Epoch of Reionization (EoR)}} \\
\textbf{Goal} &  \multicolumn{3}{c}{ \textbf{to the Epoch of Stellar buildup (ESB)}} \\
\hline
  \hline
 \textbf{Science Topics} &  Galaxy Properties  &  Clusters    & Galaxy Properties  \\
   \textbf{}           &   at the ESB  &  at the ESB      &  at the EoR \\
  \hline
    \textbf{Science Goals}    & Probe the molecular gas content & Detect gas rich clusters &  Probe the molecular gas content  \\
     {}                       & at $2.4<z < 3.4$.        & at $z\sim3$.  & at $5.8<z < 7.8$.  \\
    \hline
    \textbf{Scientific }  & High significance detection of  & Pinpointing the position of a & Tight 3 to 6$\sigma$ upper limit on the\\
    \textbf{Objectives}  & the intensity, and redshift & few hundreds of galaxy clusters &  ${\rm CO}(2-1)$ signal at $z\sim6$ plus  \\
    \textbf{}  & evolution of the ${\rm CO}(1-0)$ line & at $2.4<z < 3.4$  & estimate of the same signal.\\
    \textbf{}  & at $2.4<z < 3.4$.  & ($\sim 80\, -\, 180$ per field). & \\     
    \hline
    \textbf{Analysis methods} & Stacking of COMAP CO voxels & Selection cuts in COMAP S/N  & Masking of COMAP voxels  \\
    \textbf{  } & overlaping with one or more  & and LAEs number density are   & overlapping with LAEs followed  \\
    \textbf{} & LAEs over 4 redshift bins.  &  used to identify COMAP voxels & by: 1) stacking of COMAP voxels \\
    \textbf{} &  & bright in CO.  & (upper limit); 2) Subtraction of \\
    \textbf{} &                 &      & stacked HF ($30\,-\,34$ GHz) and \\
     \textbf{} &                 &      &  LF ($26\,-\, 30$ GHz) voxels\\
      \textbf{} &                 &      & (CO estimate). \\
  \hline
  \textbf{Analysis Section} &  \ref{subsec:Analysis.1}  &  \ref{subsec:Analysis.2}    & \ref{subsec:Analysis.4}, \ref{subsec:Analysis.5}  \\
  \textbf{Number} &    &        &   \\
  \hline
  \hline
\end{tabular}
\label{tab2}
\end{table*}

\begin{table*}
\centering
\caption{Science program and analysis methods: Secondary Goals}
\begin{tabular}{l | c | c| c}
\hline\hline
    \textbf{Overall Science } &  \multicolumn{3}{c}{\textbf{Galaxy evolution from the Epoch of Reionization (EoR)} } \\
\textbf{Goal} &  \multicolumn{3}{c}{ \textbf{to the Epoch of Stellar buildup (ESB)}} \\
\hline
  \hline
 \textbf{Science Topics} &  Galaxy clusters Properties  & Galaxy evolution from the ESB &  Galaxy Properties at the ESB\\
   \textbf{}           & at the ESB  & to the EoR     & at the ESB\\
  \hline
    \textbf{Science Goals}    & Probe the molecular gas  & Probe the evolution & Characterizing the  \\
     {}                       & clustering and velocity in & of the CO emission from  & properties of \\
     {}                       & $2.4<z < 3.4$ galaxies.  & galaxies from $z\sim3$ to $z\sim6$.  & clusters at $z\sim3$.\\
    \hline
    \textbf{Scientific }  & Probe the overall CO line & Determine how much the CO  & Determining how CO and Ly-$\alpha$ \\
    \textbf{Objectives}  & width from emission in gas & emission from low-J lines & correlate at $\sim$10 Mpc scales. \\
    \textbf{}  &  rich galaxy clusters.  & decays from $z\sim3$ to $z\sim6$. & Determining if the two lines \\
  \textbf{}  & & & trace the same galaxies. \\
    \hline
    \textbf{Analysis methods} & Probe excess signal around &  Assume that $T_{\rm CO(1-0)}\simeq T_{\rm CO(2-1)}$ at  & Probe the spatial correlation  \\
    {} & individual and stacked CO & $z\sim6$. The constrain on the CO  & between CO and LAEs by \\
    {} & bright voxels (using a 2 MHz & signal at $z\sim6$ and $z\sim3$ are & comparing the VID of  regions  \\
    {} & resolution). The stack contains & combined to probe the CO signal  & overlapping or not with LAEs. \\
    {} & only the best bright voxels & evolution. If we assume that the  & High resolution spectra (2 MHz)  \\
      {} &  oriented in frequency according & signal decays by $\sim$ 1/3 due to lower  & of bright voxels to probe the \\
        {} & to the most likely signal peak. & gas masses any  additional signal & cluster size. Potential follow up \\
        {} &  &  decay is an indication of evolution  &  of bright voxels by external \\
         {} & of LCO/MH2.  &  &surveys.\\
  \hline
  \textbf{Analysis Section} &  \ref{subsec:Analysis.3} &  \ref{subsec:Analysis.1}, \ref{subsec:Analysis.4} and \ref{subsec:Analysis.5} &   \ref{subsec:Analysis.3} and \ref{subsec:Analysis.6}\\
  \textbf{Number} &    &   &  \\
  \hline
  \hline
\end{tabular}
\label{tab3}
\end{table*}

\section{Instruments and surveys}

Two of COMAP's three fields overlap with the HETDEX coverage area and the third can potentially be observed using additional, targeted observations using the HETDEX instrument (Visible Integral-Field Replicable Unit Spectrograph; VIRUS) (\cite{hil18a};Hill, G.J. et al 2021, AJ submitted). The nominal HETDEX fill factor is 1/4.5, but further observations with VIRUS beyond HETDEX could allow one or more COMAP fields to be filled in with a fill-factor closer to unity (``full fill''). 

In this study we assume that at least one of COMAP's three fields will be observed by VIRUS with a full fill factor as follow-up observations outside of the HETDEX survey, but for the purposes of this paper, we will not distinguish between HETDEX and follow-up observations using VIRUS.

HETDEX has a higher angular resolution than COMAP and a somewhat lower frequency resolution. In our analysis, we degrade both to a common spatial and frequency resolution as shown in Table~\ref{tab:instruments} under the COMAP.H field.  

\subsection{Experimental setup for HETDEX}

The HETDEX mission proposes to trace large scale structure by mapping the positions of thousands of star forming galaxies through their Lyman alpha emission (over the redshift range $z\, =\, 1.9\, -\, 3.5$). 
This instrument will observe over two fields covering ${450\, \rm deg}^2$ with a 1/4.5 fill factor. 

HETDEX outputs include a catalog of 4/5$\sigma$ LAEs (with a $5 \sigma$ line flux limit of $3.5 \times 10^{-17} \, ({\rm erg\, cm^{-2}\, s^{-1}})$. HETDEX has a FWHM of 1.5 arcsec and a spectral resolution of R(450 nm) = 800. HETDEX sources will be carefully separated as LAEs and any interloping lines (such as OII), prior to combining it with COMAP maps. Note that while in intensity the LAEs signals should be stronger than that of the OII emission, in terms of number density the OII emitters will be more numerous than the target LAEs. 
The proposed synergy study will not be meaningfully affected by any eventual residual contamination in the LAEs catalog.

In addition to voxel statistics, the LAEs catalog, would also be useful to cross-correlate with COMAP maps \cite{2019ApJ...872..186C}. This cross correlation would only be proportional to signals originating from the same structures (redshift), hence residual OII signal will not affect this result.

\subsection{Experimental setup for COMAP Pathfinder}

The COMAP Pathfinder instrument (Lamb et al. in prep.) is a 19-feed spectrometer array fitted on a 10.4-m telescope at the Owens Valley Radio Observatory. Operating at 26--34\,GHz, this instrument is sensitive to ${\rm CO}(1-0)$ line emission from structures at $z=2.4$--3.4 and ${\rm CO}(2-1)$ from $z=6$--8. A survey of three fields using the Pathfinder began in September 2019 and is expected to run until at least 2025. It is planned to augment the Pathfinder with additional 26--34\,GHz receivers on other 10.4-m telescopes at OVRO as well as a 12--20\,GHz receiver on a prototype ngVLA antenna. The latter receiver will target ${\rm CO}(1-0)$ at $z=4$--9 (Breysse et al.\ in prep.\ ).

In Table~\ref{tab:instruments} we show COMAP Pathfinder parameters used for the proposed analyses of COMAP data making use of the HETDEX catalog (hereafter referred to as COMAP.H).

\begin{table}[ht]
\centering            
\caption{COMAP parameters assumed in this study. Values in parentheses are those prior to degradation to a common resolution with HETDEX.}            
\begin{tabular}{l  c}        
\hline\hline                 
Survey            & COMAP.H (COMAP) \\ 
\hline                 
    $T_{\rm sys}\, [{\rm K}]$    & 44    \\  
    Beam FWHM at 30 GHz (arcmin)   & 4.5          \\  
    Frequency coverage [GHz] & 26 - 34  \\
    Channel width [MHz]  & 41.5 (2.0)   \\
    Number of feeds  & 19 \\
    Sky coverage per field\, [${\rm deg}^2$] & 2.25  \\
    Number of spatial pixels       & 20 $\times$ 20  \\
    Number of frequency channels & 192 (4,096)\\
      Number of voxels per field       & 76,800 (1600,000)  \\
    Observing time per field [h] & 2,000\\
    Noise per voxel \, $[\mu {\rm K}]$ & 11.7   \\
    (used in this study) & \\
  \hline                                  
\end{tabular}
\label{tab:instruments}     
\end{table}

The thermal noise level per COMAP voxel is given by:

\begin{equation}
\label{eq:noise}
    \sigma = \frac{T_{\rm sys}}{(\tau\, \delta_{\nu})^{1/2}} = \frac{T_{\rm sys} N_{\rm  pixels}^{1/2}}{(\tau_{\rm tot}\, N_{\rm feeds} \delta_{\nu})^{1/2}} 
\end{equation}

\noindent where $T_{\rm sys}$ is the system temperature, $\tau$ is the observational time per voxel, $\tau_{\rm tot}$ is the total observational time per voxel, $\delta_{\nu}$ is the frequency resolution used in the analysis (after being degraded to match HETDEX resolution), $N_{\rm pixels}$ is the number of pixels and $N_{\rm feeds}$ is the number of feeds.

Equation \ref{eq:noise} accounts for the noise prior to any losses during the first post-processing procedure where a fraction of the total data ($f_{\rm cut}$) is removed due to filtering or flagging (see Ihle et al in prep). The noise level per voxel after the later procedure corresponds to the ``real'' noise per voxel used in the analysis with HETDEX and it is given by $\sigma_{\rm COMAP.H}$ = $\sigma\, (1 - {\rm Losses})^{-0.5}$. 
By default for this study we assume $\sigma_{\rm COMAP.H} = 11.7\, {\rm \mu K}$ which corresponds to 2000 hours of observation per field (corresponding to observing for two calendar years - with zero losses). Note that losses can be compensated by a longer observing time and that many of the analysis procedures described here would still lead to interesting results assuming a smaller integration time.  

Thermal noise in COMAP maps follows a Gaussian distribution and so the overall noise from an average (referred-to as a ``stack'') of COMAP voxels ($N_{\rm voxels}$) is:
\begin{equation}
\sigma_{\rm stack} = \sigma_{\rm COMAP.H} N_{\rm voxels}^{-1/2}.
\end{equation}

A stack of the total data in the three COMAP fields would contain $N_{\rm voxels}$ = 278784, which sets the minimum possible stack noise at $\sim 0.024\, {\rm \mu K}$ (assuming no losses).

\section{Simulations and line modelling}
\label{sec:3}
\subsection{Galaxies simulation}

Mock observational maps for both COMAP and HETDEX were made based on the TNG300 galaxy simulation. Output particles and galaxy data are available for the redshift range relevant for this study. Moreover TNG300 has an area of $(300\, {\rm Mpc})^2$ which is larger than a COMAP field size ($\sim 165\, {\rm Mpc}^2$ for CO(1-0) and $\sim (227\, {\rm Mpc})^2$ for CO(2-1)). This simulation adopts the following cosmology: $h\, =\, 0.6774$, $\sigma_8$ = 0.8159, $\Omega_{\rm M}\, =\, 0.3089$, $\Omega_{\Lambda}\, =\, 0.6911$, $\Omega_{\rm b}\, =\, 0.0486$ and $n_s\, =\, 0.9667$. 

The TNG300 simulation was post-processed using different prescriptions for the CO and LAE emission to explore the possible range of signals. The CO line modelling was based on Baumschlager et al. (in prep.) and the LAE modeling was done specifically for this study. 

\subsection{CO line modelling}

\begin{figure}
    \includegraphics[width=\columnwidth]{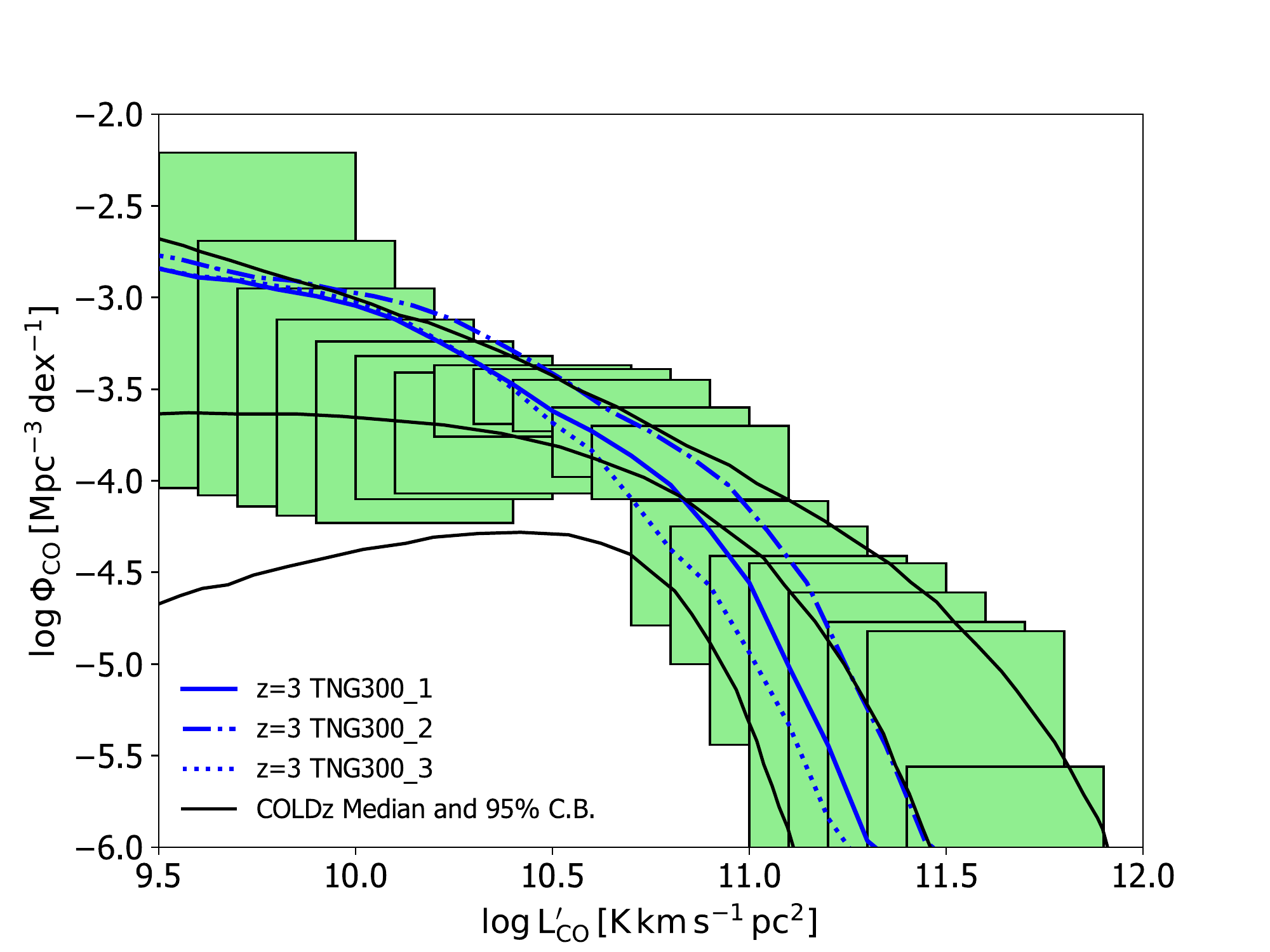}
\caption{Simulated ${\rm CO}(1-0)$ Luminosity Function at $z=3$ (models TNG300$\_1$, TNG300$\_2$ and TNG300$\_3$ from Baumschlager et al. (in prep)) are shown in blue.
The green boxes mark the constraints from COLDz data centered at $z=2.4$. Black lines denote the median and 90\% confidence boundary of the COLDz data.}    
\label{Fig:1}
\end{figure}

Mock CO maps with COMAP.H resolution were produced
from three CO simulations described in (Baumschlager et
al. (in prep)) as TNG300$\_1$, TNG300$\_2$ and TNG300$\_3$, hereafter referred to respectively
the medium, optimistic and pessimistic model. 
The Baumschlager CO simulations were obtained by post-processing simulations of galaxy formation with semi-analytical methods to derive molecular gas masses, and to convert the latter into CO luminosities. The three CO models selected for this study, take as base the TNG300 simulation (galaxies and particles data) and were chosen based on their fit to observational constraints from the COLDz survey.  This study was made by directly using the simulation outputs from the Baumschlager et al. study.  CO luminosity functions (LFs) derived from the simulations at ${z=3}$ are shown in Figure~\ref{Fig:1} in comparison with current constraints from the COLDz galaxy survey. CO prescriptions for the full redshift range covered by COMAP for the ${\rm CO}(1-0)$ and ${\rm CO}(2-1)$ lines are available in the same study.

The intensity of the CO emission can be derived from the line luminosities using:

\begin{equation}
    I^J_{\rm CO}(\nu)=\int^{M_{\rm min}} _{M_{\rm max}} dM \frac{dn}{dM} \frac{L^{\rm J}_{\rm CO}}{4 \pi D^2_L} y\left(z(\nu,J)\right) D^2_{\rm A},
\end{equation}

\noindent where $y\left(z(\nu,J)\right)=d\chi/d\nu=\lambda_{\rm CO}(1+z)^2/H(z)$ and the integration was done over halo mass. The CO brightness temperature is given by:

\begin{equation}
    T^{\rm J}_{\rm CO} = \frac{I^{J}_{\rm CO}(\nu)\, c^2 }{2\nu^2\, k_B},
\end{equation}

\noindent where $k_B$ is the Boltzmann constant and c is the light velocity.

\subsection{The CO line width}
\label{subsec:EW}
Due to the large velocity of the emitting gas the observed CO line will be broad. The CO line width is correlated with the mass of the halo containing the galaxy, the turbulence of the gas and the galaxy shape and inclination.

As the halo increases in size so does on average its molecular gas content and overall line luminosity. A positive correlation is commonly observed between the CO(1-0) line full width at half maximum (FWHM) and luminosity \citep{2019ApJ...882..136A}. 

\begin{figure}
    \includegraphics[width=\columnwidth]{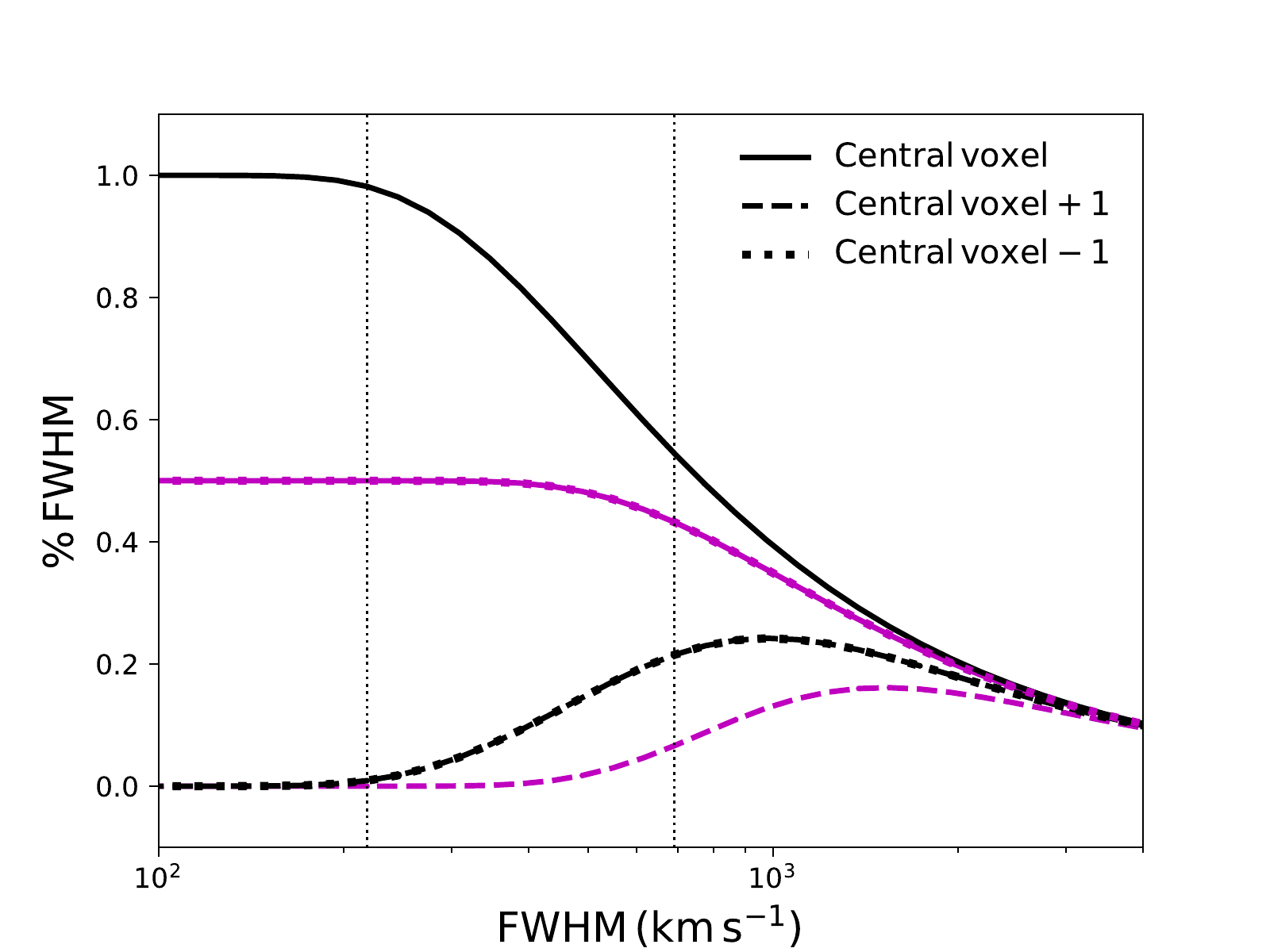} 
\caption{Fraction of the galaxy line width observed in the central voxel and in adjacent voxels in the spectral direction. Black lines assume that the galaxy is located in the center of the voxel and magenta lines assume that the galaxy is located right at the beginning of the voxel. Black dotted vertical lines denote the average FWHM of galaxies with $L^{\prime}_{\rm CO}=10^{\rm 10}\, {\rm K\, km\, s^{-1}\, pc^2}$ (left line) and $L^{\prime}_{\rm CO}= 10^{\rm 11}\, {\rm K\, km\, s^{-1}\, pc^2}$ (right line).  This figure shows above which FWHM values the signal of a galaxy will be observed over several COMAP voxels.}    
\label{Fig:2}
\end{figure}

Given the frequency resolution of the COMAP.H survey, the CO line will not be resolved for the large majority of galaxies (moreover, COMAP spatial resolution does not make it possible to separate the signal from individual sources). However bright emitters ($L'_{\rm CO}\gtrsim 10^{10}\, {\rm K\, km\, s^{-1}\, pc^{-2}}$) will often be observed over several COMAP voxels. Moreover, some galaxies will be located at the edge of a voxel which increases the probability of the signal being spread over multiple voxels. Figure~\ref{Fig:2} illustrates the need to account for the CO line equivalent width (EW).

\begin{figure}
\includegraphics[width=\columnwidth]{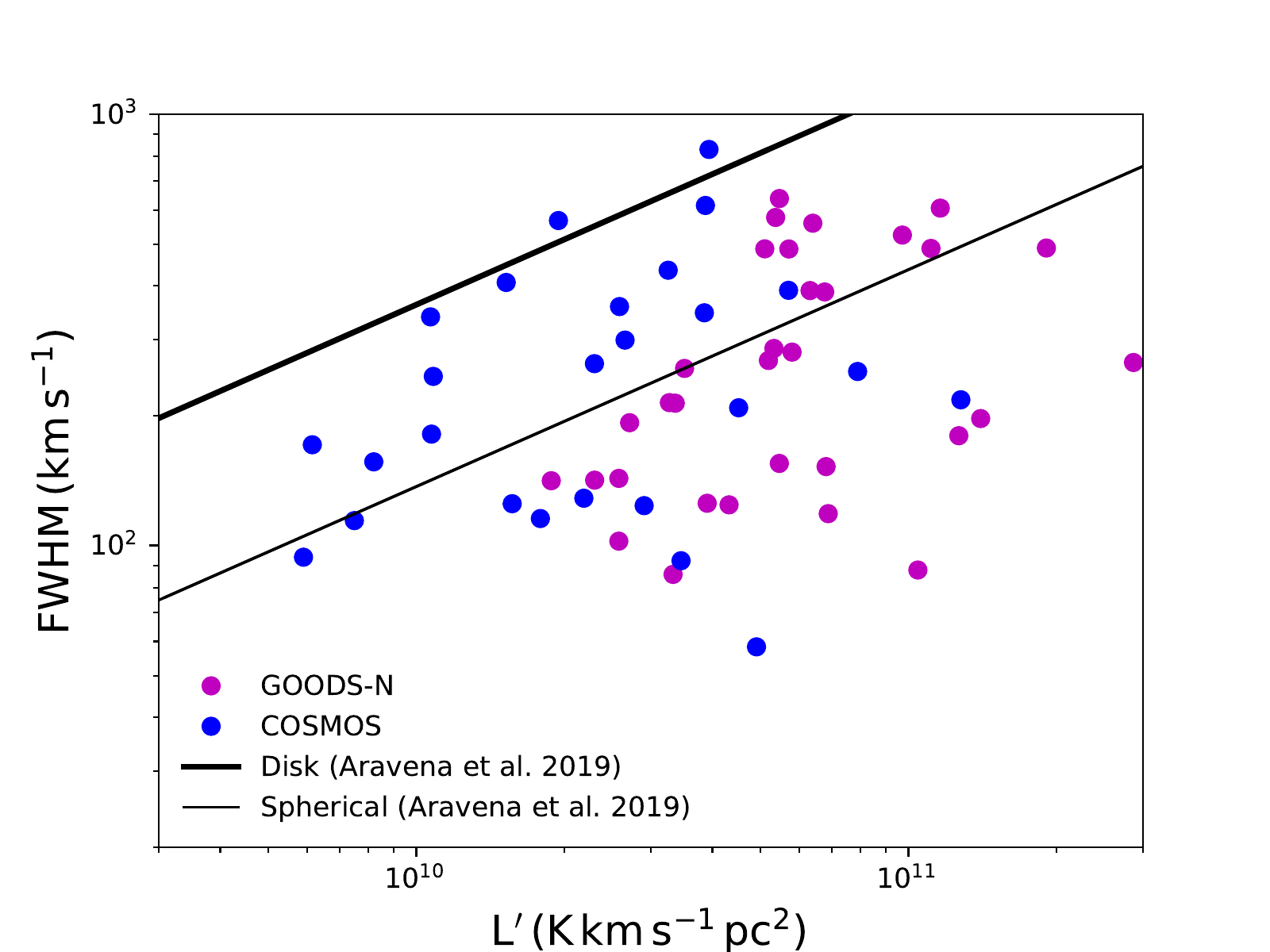}
\caption{FWHM as a function of CO luminosity. Dots show the FWHM of COLDz galaxies (in both the deeper COSMOS and wider GOODS-N fields) and solid thick and black lines show the expected relation for respectively disk and spherical like galaxies \citep{2019ApJ...882..136A}.}    
\label{Fig:3}
\end{figure}

The effect of the CO EW on power spectra derived from COMAP observations was explored in detail in \cite{2021arXiv210411171C}.
In this study, we account for the FWHM smoothing of the observed signal (in frequency) by modeling the CO line width for the simulated galaxies following the \cite{2019ApJ...882..136A} prescription for spherical sources. The FWHM is given by:

\begin{equation}
\left(\frac{\rm FWHM}{2.35}\right)^2=\frac{L^{\prime}_{\rm CO}}{\rm C}\left(\frac{\alpha_{\rm CO}\, G}{\rm R} \right), 
\label{eq:FWHM}
\end{equation}

\noindent with the FWHM in ${\rm km\ s^{-1}}$, ${\rm R}$ is the source radius and G the gravitational constant. To the FWHM derived from equation \ref{eq:FWHM} we added a random 0.2 dex uncertainty. 

We adopt the values ${\rm R\, =\, 2\, kpc}$, ${\rm C\, =\, 5}$ and $\alpha_{\rm CO}=0.8$ as in the \cite{2019ApJ...882..136A} study. We justify our choice of $\alpha_{\rm CO}$ by noting that the galaxies with FWHM values well above our survey resolution correspond to very luminous sources which often have starburst like $\alpha_{\rm CO}$ values. Moreover, as can be observed in Figure~\ref{Fig:3} this fitting function provides a good fit compared to the FWHM of COLDz survey galaxies.

We assume a log-normal scatter of 0.2 in the median relation consistent with that of COLDz. Only the signal from galaxies with CO luminosities above the COLDz survey sensitivity limit is expected to be observed over several voxels. 
Towards high redshift ($z\sim6$ or more) as the typical properties of galaxies evolve and especially in the presence of strong turbulence the FWHM of the line will increase. As this turbulence also boosts the luminosity of the line it is expected to conserve the overall correlation between the line FWHM and luminosity.  

We explore the impact of the CO EW on the observed spatial correlations between CO emission and LAE in Section \ref{subsec:Analysis.3}. In the same section, we also explore the possibility to measure the CO EW directly from COMAP data. We note that the native COMAP frequency resolution is 2 MHz which can aid the analysis of bright CO voxels.
In a following study, we plan to explore the relation between FWHM from COMAP CO emitters versus HETDEX LAEs directly from the observational data.

Overall, there is significant potential to probe galaxy emission line EWs with LIM surveys and COMAP should be the first LIM mission that can pursue this goal due to its high frequency resolution. 

\subsection{Mock simulations of LAE}
 
We adopt the same base galaxy simulation to generate mock observations for HETDEX LAEs as was used for the COMAP CO maps. 

Following \cite{1998ARA&A..36..189K} and \cite{2017arXiv171109902S}, the intrinsic Lyman-$\alpha$ luminosity should scale with the galaxy star formation rate (SFR) as 
\begin{equation}
    L_{Ly\alpha}^{\rm int}\, {\rm [erg\, s^{-1}]} = 1.1 \times 10^{42}\, SFR {\rm [M_{\odot}\, yr^{-1}]}. 
\end{equation}
Here, we used the SFR from the galaxies in the TNG300 simulation to derive the intrinsic line luminosities. 
Lyman-$\alpha$ photons are easily absorbed in a galaxy's neutral gas which increases their probability of being scattered out of the line of sight or absorbed by dust grains. 

The Ly$\alpha$ luminosity corrected for dust absorption is therefore much lower than the galaxy intrinsic luminosity and can be obtained using:
\begin{equation}
     L_{Ly\alpha}^{\rm dust} =  L_{Ly\alpha}^{\rm int}f_{\rm dust} 
\end{equation}
where $f_{\rm dust}$ is the fraction of photons that escape the galaxy without being absorbed by dust which is a function of several galaxy properties such as its dust content and the burstiness of the galaxy star formation activity. Most observed galaxies have very low escape fraction of Lyman alpha photons of the order of 1\% or less. A small fraction of galaxies of the order of $10\%$ are Lyman alpha emitters (LAEs) and have escape fractions that can reach 100\% (as calculated using the ratio between the H$\alpha$ and Ly$\alpha$ lines) \citep{2009A&A...506L...1A}.
However, the average escape fraction of LAEs at $z\sim2-3$ is closer to (5- 30)$\%$ (depending on the galaxy sample and Lyman alpha flux limit) \citep{2019A&A...623A.157S}. Observations show that LAE are associated with medium size, dust-poor galaxies with large specific star formation rates. 
Galaxies can change between the main sequence and starbusting phase causing large time fluctuations in the Lyman alpha signal which is particularly high during outbursts.

A precise calculation of the fraction of Lyman alpha photons that escape a galaxy without being absorbed by dust or scattered out of the line of sight, for each individual galaxy, requires adopting a specific shape for the intrinsic Lyman-$\alpha$ line, accounting for the galaxy type and inclination and using a radiative transfer code to simulate the random walk of Lyman alpha photons in the galaxy gas and the observed line shape (see eg. \cite{2018A&A...614A..31B}). Such a complex estimation is model dependent and beyond the scope of this study. Given that our goal is only to make a rough estimation of which galaxies would be observable by HETDEX we adopt the following three prescriptions ($i$, $ii$ and $iii$) to account for the dependence of the observed signal on the galaxy dust to gas ratio $({\rm DGR}=M_{\rm dust}/M_{\rm n})$. The TNG300 simulation does not provide dust masses and so we use metallicity as a proxy for dust mass and determine DGR, as in Baumschlager et al., using the galaxy neutral gas masses ($M_{\rm n}$) and metallicity ($Z$). Note that observed LAEs usually have very low metallicities ($Z\, <\, 0.4\, {Z_{\odot}}$ \cite{2012ApJ...745...12N}. We model the fraction of Lyman alpha photons which are not absorbed by dust as: 

\begin{eqnarray}
\label{eq:fdust}
    f^i_{\rm dust} &=& 10^{- {\rm DGR/DGR}_{\rm MW}}, \nonumber \\ 
    f^{ii}_{\rm dust} &=& 10^{-1.3 {\rm DGR/DGR}_{\rm MW}},
    \nonumber \\ 
    f^{iii}_{\rm dust} &=& 10^{-1.6 {\rm DGR/DGR}_{\rm MW}}.
\end{eqnarray}

\noindent Where it is assumed that $f_{\rm dust}$ scales with the galaxy dust content as a power law, similar to what is used to correct for galaxy reddening. The adopted extinction coefficients ($i=1$, $ii=1.3$ and $iii=1.6$) imply a softer dependence of the escape fraction on the DGR compared to the dependence of $f_{\rm dust}$ on galaxy reddening which is observationally predicted to be much steeper ($\sim\, 2.7$). These prescriptions result in LAEs with ${ f_{\rm dust}[\%]=(i,43);(ii,35);(iii,28)}$ (for luminosities above $10^{42}\, {\rm erg\, s^{-1}}$) which are comparable with observational measurements at similar redshifts which estimate order of $30\%$ escape fraction \citep{2014ApJ...787....9W,2011ApJ...736...31B}.  
We limit the overall $f_{\rm dust}$ between zero and one and adopt prescription $i$ as our fiducial model.
These corrections ensure that very dusty galaxies are not LAEs. 
We follow by randomly selecting only a fraction of the galaxy population as LAEs; in practice we introduce a factor $f_{\rm LAE}=1/0$ which denotes the galaxy being or not a LAE. The fraction of LAEs changes with luminosity bin so that the final Lyman alpha luminosity function fits current observations; a similar exercise was presented in \cite{2019MNRAS.487.5070M}. In the later step, we are introducing a duty cycle in the time period in which a galaxy is observed as a LAE; in reality, the probability of a galaxy being observed as a LAE depends on galaxy properties such as burstiness of the SFR. The overall fraction of Lyman alpha photons that escape a galaxy and are detectable by observations of LAEs is $f_{\rm esc}=f_{\rm dust}\times f_{\rm LAE}$.  

The observed Lyman alpha luminosity of a galaxy is then given by:
\begin{equation}
     L_{Ly\alpha}^{\rm obs} =  L_{Ly\alpha}^{\rm dust}f_{\rm LAE}. 
\end{equation}

\begin{figure}
\centering
    \includegraphics[width=\columnwidth]{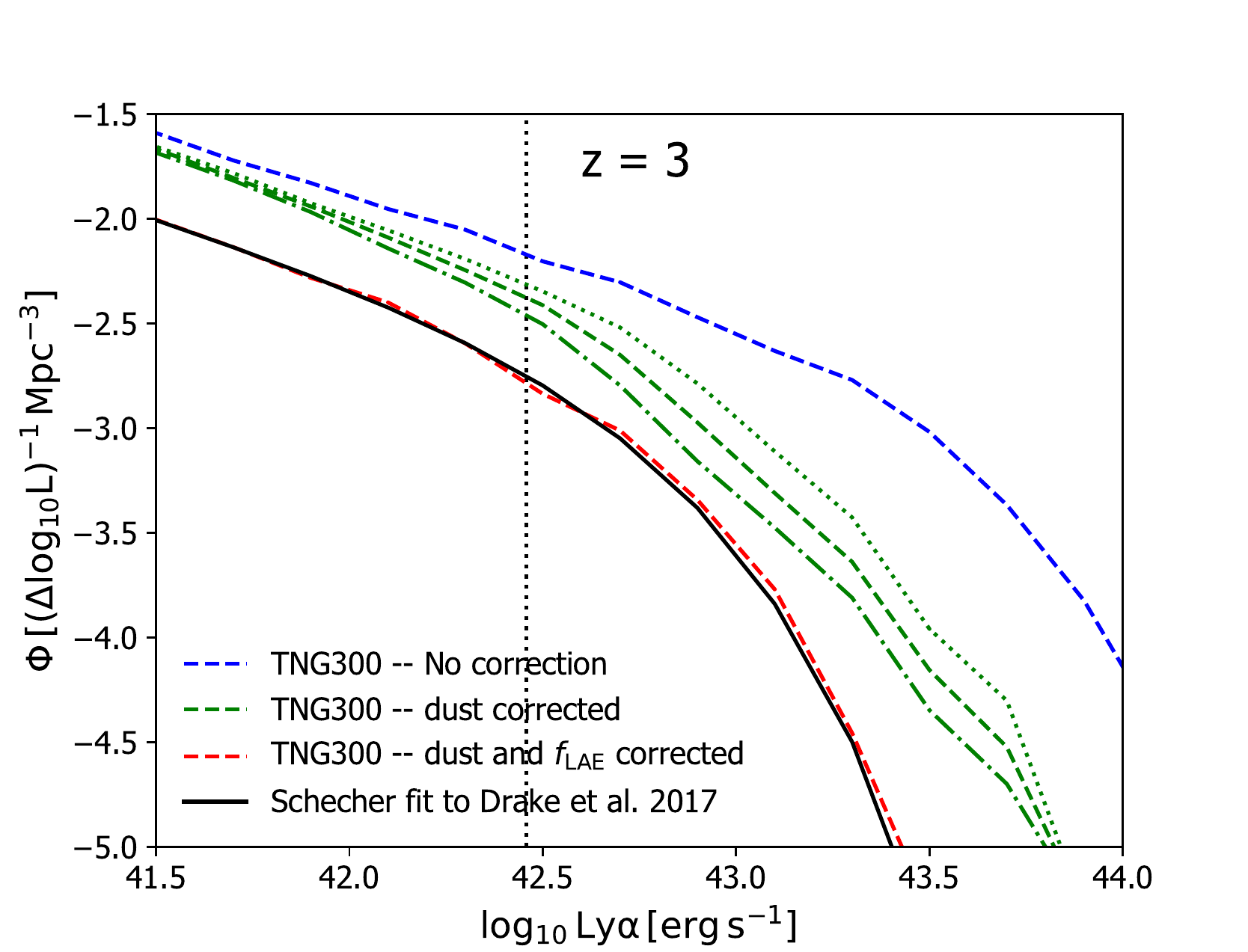}
\caption{Lyman alpha Luminosity Functions obtained from the galaxy catalog in the Illustris TNG simulation at $z\sim3$. Green dotted, dashed and dashed dotted lines correspond respectively to dust correction prescription $i$, $ii$ and $iii$ (the correlation between CO and LAE is higher for prescription 2).The red dashed line shows the final LF from LAE after being corrected for dust extinction and scattering in the galaxy (this LF is the approximately the same for the three dust prescriptions).   }       
\label{Fig:4}
\end{figure}

This method to determine LAEs insures that our simulation matches the relatively narrow observational constraints and that the LAE emission dependence on dust content is accounted for. This latter point is an essential requirement for predicting the correlation of LAEs with CO emission.
Figure~\ref{Fig:4} shows the Lyman-alpha LF before and after being corrected. 

LAEs are often observed with an offset to the galaxy systemic velocity. We correct the position of the simulated LAEs for velocity offsets as a function of galaxy SFR using the relation found in the HETDEX pilot survey \citep{2014ApJ...791....3S}. We explore the uncertainty of this relation and find that only $0.01\, -\, 0.05\%$ of the LAEs are "observed" in a different voxel due to this correction. Consequently LAEs velocity offsets will not affect synergies between COMAP and HETDEX.

\section{Data analysis: Signal separation}
\label{sec:Analysis}

A key advantage of the Line Intensity Mapping technique is that it makes it possible to detect signals close to or even below the thermal noise level. In the case of COMAP the median CO emission in a voxel should be of the order of $T_{\rm CO}^{\rm total}\,  \sim\, 1 \mu {\rm K}$ although highly variable from voxel to voxel. The COMAP survey considered here has a total noise level around an order of magnitude higher than the target signal in a single voxel, ($\sigma_{\rm voxel}\, =\, 11.7 \mu {\rm K}$). 

Various analyses can be performed on the COMAP data alone and in combination with other data-sets such as that from HETDEX. In terms of power spectrum analyses, these include the estimation of the full 3-D auto-power spectra (Chung et al. and Ihle et al. in prep) as well as the cross-power spectra between COMAP and HETDEX. This will likely be performed on halves (4 GHz slices) of the data in order to check for cosmic evolution of the signal. A high significance detection of the total CO signal (dominated by the low redshift emission) and its spatial fluctuations over a range of scales (with maximum sensitivity at $k \sim 0.1$ Mpc$^{-1}$) should be in the reach of COMAP after a few years of integration time. 

Considering voxel-level analyses, the COMAP voxels can be stacked (Ihle et al in prep.) in order to make a detection and constrain the average signal level. A stack of all voxels in a COMAP field translates to a noise level of $\sigma_{\rm stack}\sim 0.042\, {\rm \mu K}$. Given the expected high significance of a CO detection from stacking, it should be possible to divide the complete COMAP data cube into a series of frequency/redshift slices and examine the redshift evolution of the average signal level. The COMAP voxels can also be  analyzed using information from other data-sets and the various ways in which this can be done using HETDEX data is the focus of this work. 

At large scales (above a few Mpc), the spatial distribution of the CO emission traces that of the cosmic web filaments. Hence, according to our simulations, $10-20\%$ of COMAP voxels will contain most of the CO signal (more than $80\%$). Given that the target CO emission is inhomogenous a small fraction of the voxels (maximum a few $\%$) in a map will be above the average noise level, we hereafter refer to them as ``bright voxels'', see Table~\ref{tab:Bright voxels definition}. 

\begin{table*}
\centering            
\caption{Different definitions of ``bright voxels'' adopted in this study. }            
\begin{tabular}{l  c }        
\hline\hline                 
Type of voxel        & Definition  \\ 
\hline 
    Bright voxel  & COMAP voxel with an observed signal above the average noise level.      \\  
    HETDEX bright voxel & COMAP voxel overlapping with at least one LAE detected by HETDEX.  \\ 
    COMAP bright voxel & COMAP voxels identified as rich in CO emission with the selection cuts outlined in Section~\ref{subsec:Analysis.1}. \\
  \hline                                  
\end{tabular}
\label{tab:Bright voxels definition}     
\end{table*}

The accurate identification of CO bright voxels and the differentiation between the signal contributed by the two CO lines cannot easily be done with COMAP data alone. Bright voxels will have a signal-to-noise of order of a few and so will rarely stand out from noise peaks.
Moreover, the signals from the two CO target lines are poorly constrained from observations/simulations and the low-z signal ${\rm CO}(1-0)$ is expected to be of order of a few to one magnitude higher than the high-$z$ signal ${\rm CO}(2-1)$. Consequently, isolating the ${\rm CO(2-1)}$ line requires additional external information. 

LAE galaxies detected by HETDEX will trace the same large scale structure as COMAP (${\rm CO}(1-0)$ line), hence the position and number density of the LAEs will often be correlated with bright CO emission at a large enough scale.
CO and LAE emitters will be positively correlated with very over-dense regions and uncorrelated with voids. However, LAEs will rarely be bright in CO since the former emission line is anti-correlated with dust content and the latter is positively correlated. Hence, low density gas filaments will be bright in LAE or CO depending on their dust content, Table~\ref{tab:Bright voxels} shows a schematic description of the structures expected to be bright/faint in CO brightness or LAE number density. The voxel size used to analyze COMAP and HETDEX data is large enough to contain a mixture of several of the ``regions'' shown in the Table. Moreover, COMAP maps are noisy, hence, the classification of COMAP as corresponding to specific ``regions'' cannot be naively applied to observed data without more sophisticated analyses.

\begin{table}
\centering            
\caption{Classifying the dust content and large scale structure, in a sky area corresponding to a voxel size, based on the combined CO and LAEs signals. }            
\begin{tabular}{l  c c }        
\hline\hline                 
Survey        & LAE bright  & LAE faint  \\ 
\hline                 
    CO Bright & Knots/ galaxy cluster  & Filaments \\
              & Mixed dust regions     & High dust  \\
\hline 
    CO Faint  & Filament  & Void       \\  
    & Low dust &  Low dust   \\ 
  \hline                                  
\end{tabular}
\label{tab:Bright voxels}     
\end{table}

At low redshift ($z\sim3$), the COMAP team plans to use HETDEX LAE data to increase the detection significance of ${\rm CO(1 - 0)}$ emission and to pinpoint the location of the brightest CO clusters. We will hereafter refer to COMAP voxels that overlap with at least one LAE detected by HETDEX as ``HETDEX bright voxels'', see Table~\ref{tab:Bright voxels definition}.

Galaxy clusters and protoclusters contain a large number of galaxies, hence they are likely to contain both LAEs and CO bright galaxies. 
Moreover, given that the LAEs will only correlate with ${\rm CO}(1-0)$ emission, the position of the LAEs can be used to partition the emission from the two CO lines. We will explore how to use the external low-redshift tracer to isolate and at least put an upper limit on the ${\rm CO}(2-1)$ emission at high-redshift.

The following analyses were optimized based on a set of CO and LAE models however the overall methods are robust and should be valid for the more complex ``real'' signals. Moreover, the analyses can be extended to include additional signal models. The validity of these methods can be easily tested/confirmed with observational data. 

Unless stated otherwise, the results shown in this section were computed assuming the COMAP.H configuration, one field and an integration time such that the COMAP noise level reaches 11.7 $\mu {\rm K}$ (two to five years). We assume that HETDEX would survey with full fill coverage over one to  three of COMAP fields. The significance (${\rm S/N}$) of the detection scales with the square root of the number of fields. By default the data shown in this section were based on the use of model TNG300$\_2$ for CO and on prescription 1 for LAEs (our results are little affected by the later choice). These are the most optimistic models adopted in this paper however all the CO models we consider have a conservative low brightness compared to most other literature models.

\begin{figure*}
\centering
\includegraphics[width=0.45\linewidth]{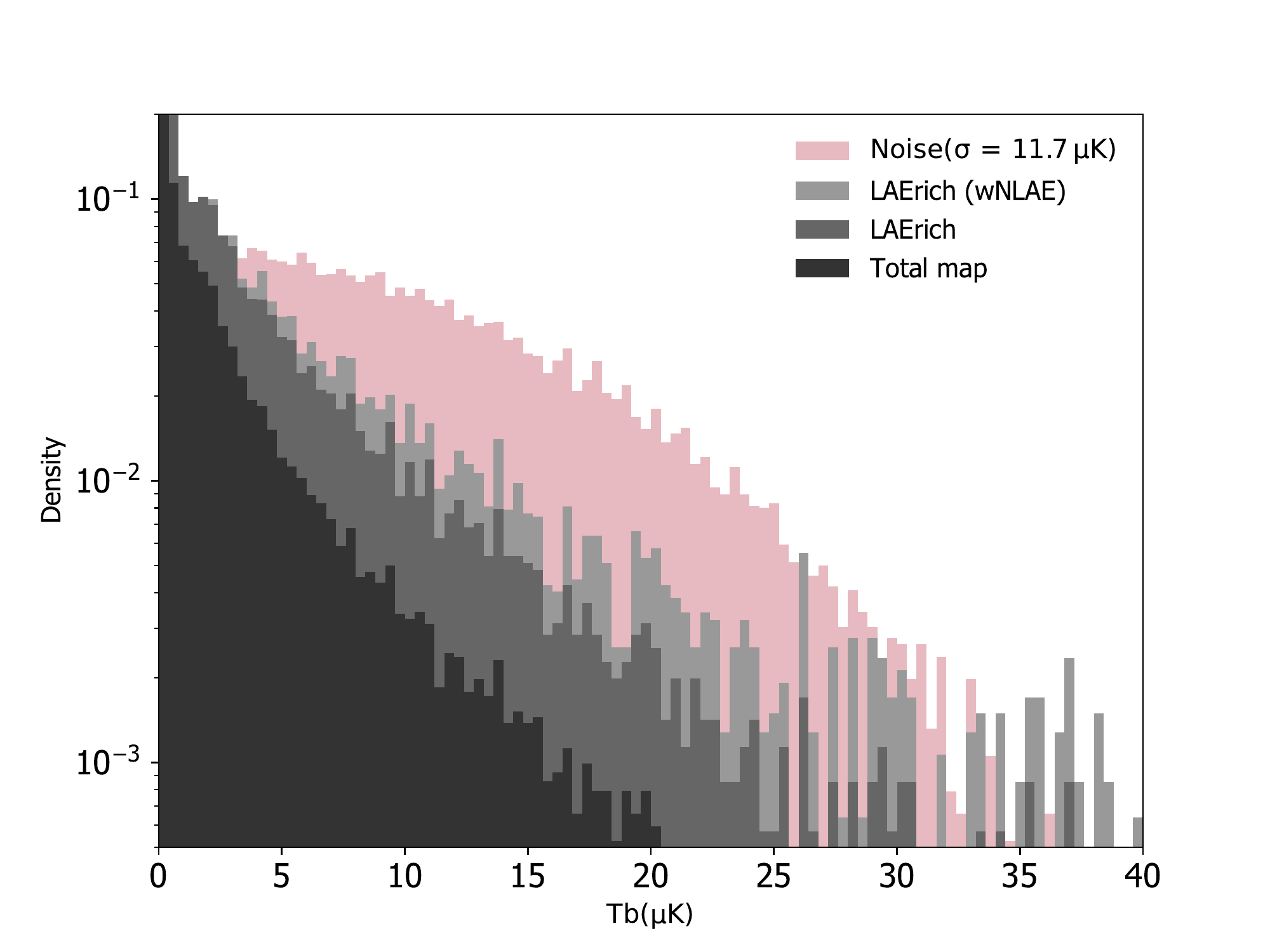}
\quad
\includegraphics[width=0.45\linewidth]{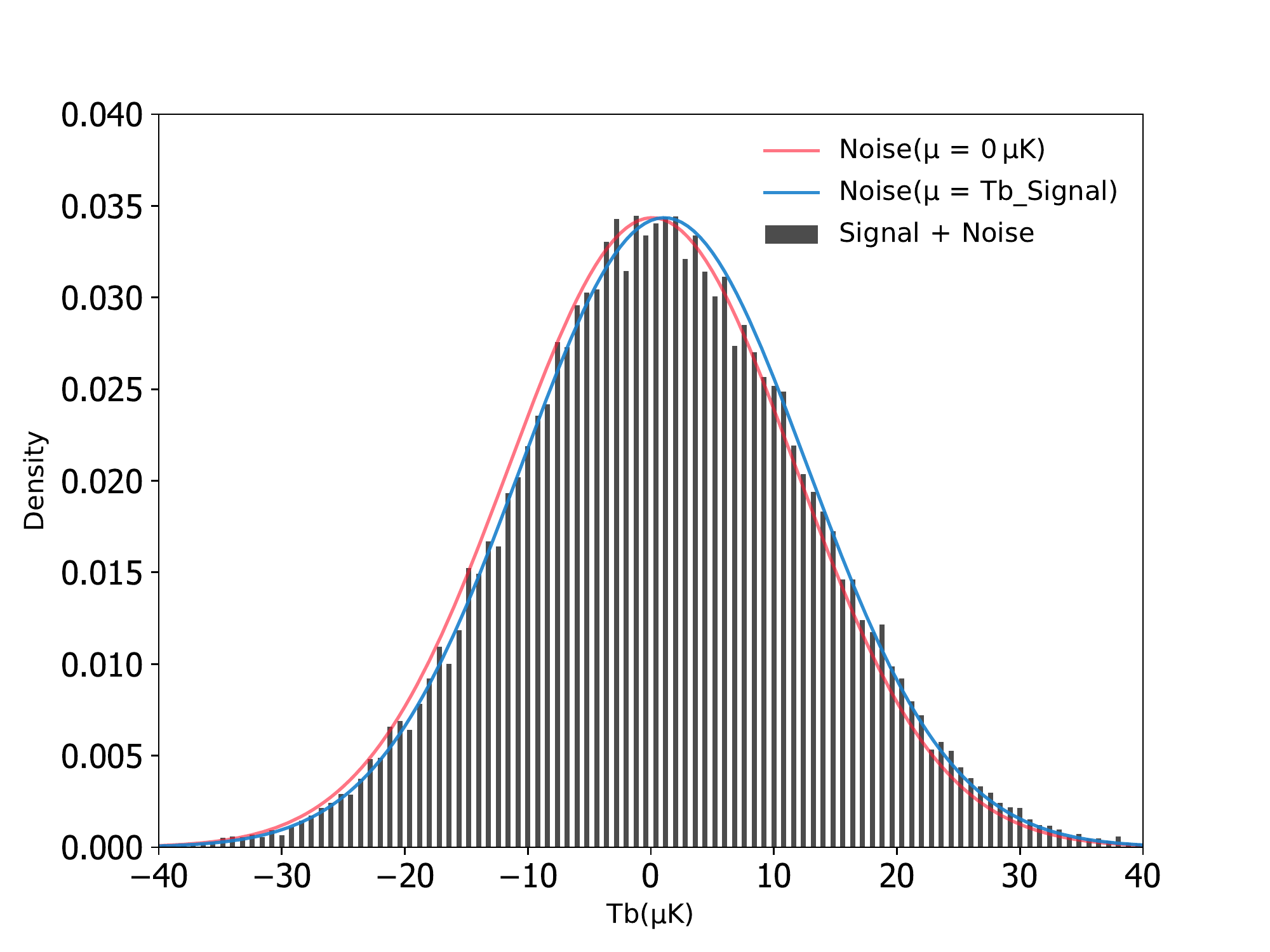}
\caption{VID of the CO signal versus noise. The left panel compares the simulated VID of the total CO emission given by the sum of the ${\rm CO}(1-0)$ and ${\rm CO}(2-1)$ signals (grey bars) with a random realization of the noise (pink bars), showing that the overall signal is buried in the noise. The left panel also shows that the VID of the COMAP voxels overlapping with LAEs (LAErich) is closer to the noise level compared to the total map. In addition, the VID of COMAP voxels overlapping with LAEs weighted by the number of LAEs (LAErich - wNLAE) provides a clear signal detection, with a much larger density of voxels with ${\rm Tb\sim 25\, \mu K}$ than the noise distribution.  The right panel compares the simulated VID of the CO signal plus noise with theoretical noise distribution curves centered at ${\rm Tb} =\, 0\, \mu {\rm K}$ (pink line) and at Tb = Tb(CO Signal) (blue line); this plot shows that the CO emission causes a ``shift'' in the overall signal from noise only, towards brighter luminosities.}
\label{Fig:5} 
\end{figure*} 

Figure~\ref{Fig:5} gives an overview of what we can expect to get from COMAP data (in terms of voxel statistics) without using the HETDEX data. As the right panel of this figure illustrates the VID from COMAP data alone will have an excess signal above ${\rm Tb}\sim 25\, \mu {\rm K}$ relative to that of Gaussian noise. therefore COMAP data alone is good enough to get a reasonable detection of the overall CO emission as long as the observational noise is reasonably Gaussian as is expected. The methodologies described in this section show what extra information we can get from the COMAP data-set by combining it with the HETDEX survey LAE catalog.

\subsection{Total CO emission at $z\sim3$: COMAP + HETDEX}
\label{subsec:Analysis.1}
COMAP will detect ${\rm CO}(1-0)$ line emission originating from the same large-scale-structures as HETDEX LAEs, hence, the two signals will be positively correlated at large enough scales. COMAP.H voxels are relatively large ($\sim 6.4\, {\rm Mpc}^3$ ) and so we expect that voxels with a large number of LAEs correspond to cluster regions which contains relatively high CO emission.

The HETDEX mission will detect order of 20,000 LAEs (with a ${\rm S/N}> 5\, \sigma$) per COMAP.H field.

Using our simulations we explored the sensitivity of a CO measurement obtained by stacking HETDEX bright voxels (COMAP voxels overlapping with at least one LAE detected by HETDEX with a significance $> 5\,\sigma$) over a 2 GHz frequency bin  and found that: 

\begin{itemize}
\item HETDEX $5\,\sigma$ LAE will be spread over $\sim 17-18\%$ of COMAP voxels. 
\item HETDEX bright voxels overlap with $\sim 55\%$ of the total CO emission (largely dominated by the low-z CO signal).
\item HETDEX bright voxels have a median CO brightness of  $T_{\rm CO}\sim 2-3\, {\rm \mu K}$ compared to ${\rm \sim1 \mu K}$ for median of all voxels.
\item The significance of the stack is of the order of $20\, \sigma$ (for one field). The significance of HETDEX bright voxels is 2 to 3 times higher than a stack of all the voxels in the map). The left panel in Figure~\ref{Fig:5} illustrates this effect by comparing the expected COMAP noise with the VID of the total map, the VID of the voxels overlapping with LAEs (referred to as LAErich) and a VID of the voxels overlapping with LAEs weighted by the number of LAEs in a voxel (LAErich - wNLAE). The VID of the weighted LAErich voxels adds an extra factor of two to the significance of a stack.
\end{itemize}
If as expected, the S/N of the CO detection (over a 4 GHz map) is of the order of several sigmas the science analysis can be made over smaller redshift bins to better constrain the redshift evolution of the CO signal and its spatial fluctuations.

\subsection{Bright CO voxels at z = 3: COMAP + HETDEX}
\label{subsec:Analysis.2}

COMAP maps will be noise dominated. However, given that the signal is highly inhomogeneous, a fraction of the COMAP voxels will have a signal a few sigma above the noise. The distribution of peaks of CO emission can be probed by assuming that it is responsible for the deviation from the Gaussian like distribution expected for the noise peaks.

Given the expected large CO emission in HETDEX bright voxels we explored cuts in HETDEX LAEs number density and in CO emission that can be used to discriminate between bright CO emission and noise in COMAP maps. We will hereafter refer to the voxels identified with this method as ``COMAP bright voxels'' (see Table~\ref{tab:Bright voxels definition}).

\begin{figure*}
\includegraphics[width=18.4cm]{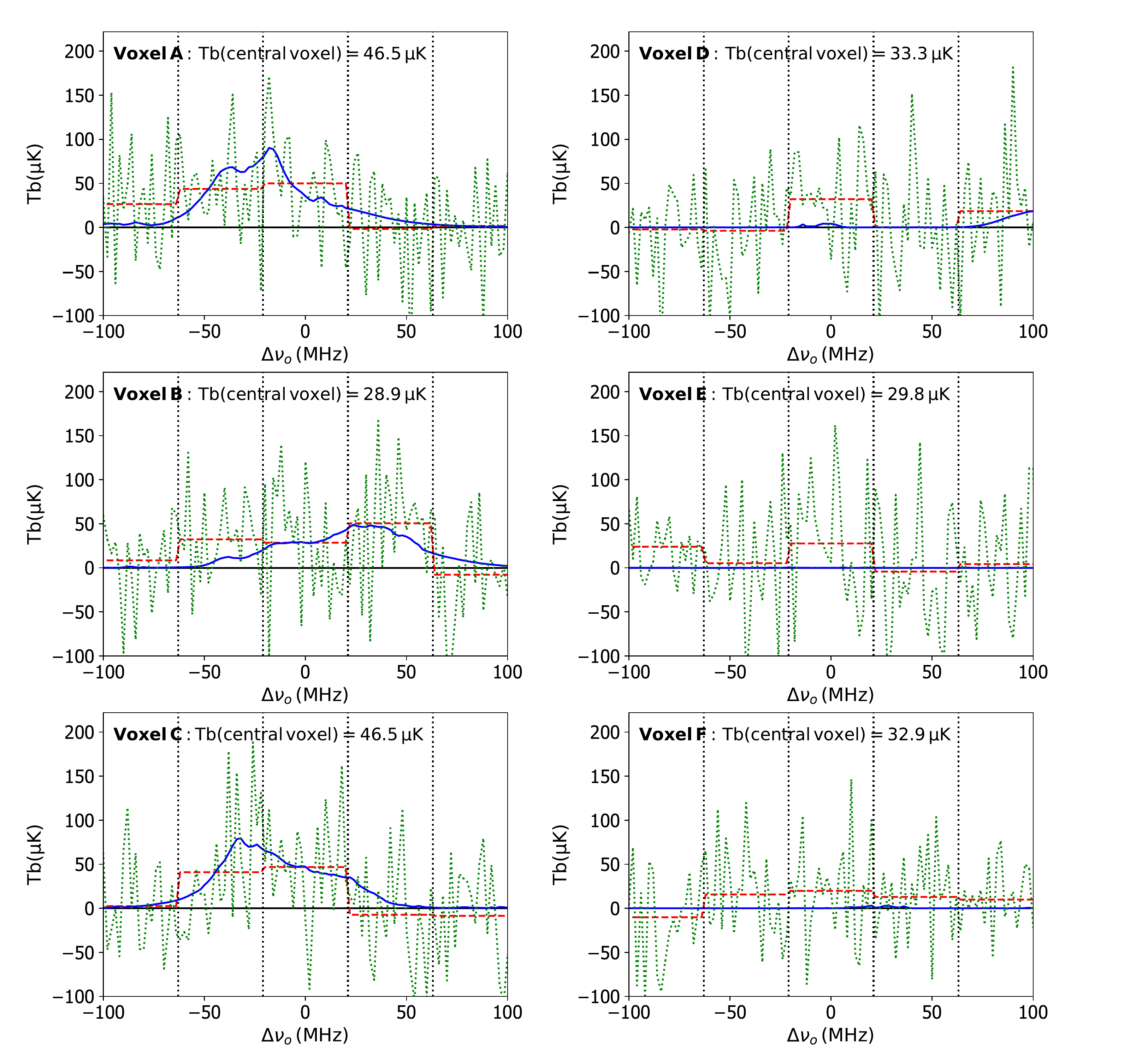}
\caption{CO brightness temperature around a candidate to COMAP bright voxel, in the spectral direction, with a low noise obtained using a 2 MHz spectral resolution, for six of the brightest observed voxels. The central voxel was identified following the description from Section~\ref{subsec:Analysis.2}. The green dotted line and red dashed line show the observed signal (including noise) with respective the high and low frequency resolution, the blue solid line shows only the signal. Black vertical lines indicate the limits of a voxel with the COMAP.H spectral resolution. The plots on the left correspond to actual COMAP bright voxels and the plots on the right correspond to noise fluctuations which would pass our criteria to identify COMAP bright voxels but that could be rejected after visual examination; we note that we expect that only a small percentage ($<5\%$) of misidentified COMAP bright voxels prior to the visual analysis.}    
\label{Fig:6}
\end{figure*}

We found that:
\begin{itemize}
\item The probability of a HETDEX bright voxel being CO bright increases with the number of LAEs in the voxel. There is also a positive correlation between CO brightness and LAEs flux but it is not as significant.
\item Voxels with three or more LAEs and with an observed signal above the average noise ($T_{\rm CO}^{\rm obs}/N_{\rm av}>1.5$) will in most cases ($> 90\, \%$) pinpoint regions with strong CO emission ($T_{\rm CO}^{\rm True}/N_{\rm av}>1$). Moreover, in these voxels the observed emission will usually be signal dominated ($T_{\rm CO}^{\rm True}/N_{\rm True}>1$).

\item Given that one of the criteria to identify COMAP bright voxels is that the voxel has an observed signal $>1.5$ times the expected noise level, these voxels will have a noise level slightly above the average.

\item A small fraction of the voxels detected with this method will still be noise dominated (our simulations predict this fraction to be $\sim 10\%$). Nevertheless, even these noisy voxels should still correspond to regions relatively bright in CO.

\item Our simulations predict a detection of 90 to 200 bright CO voxels (per field) with this method (the uncertainty accounts for the different CO and LAEs models). 

\item We analyzed the sources of the emission in the bright voxels identified using this method and found that typically $50$--80$\%$ of the emission is due to a single galaxy or to two galaxies.

\item Due to the relatively large dust content of CO bright galaxies, we found that the brightest CO emitters are located in dust rich regions with a low number density of LAEs (as would be expected according to Table~\ref{tab:Bright voxels}). Hence, the brightest CO emitters will often not be identified as bright CO voxels using this method.

\item We found that COMAP voxels that are adjacent in the spectral direction to bright COMAP voxels, often have a CO brightness well above the average (3 to 7 $\mu K$) and an average or below average noise level.

\item In our simulations bright CO voxel candidates with a bright emission from adjacent voxels (in the spectral direction) usually correspond to real CO bright voxels (as against noise peaks), see Figure~\ref{Fig:6}.

\item The signal in bright CO voxels, can be re-scaled to a higher frequency resolution (up to 2 MHz). This can help to distinguish between real CO signals and noise peaks given that in the case of a real signal, due to clustering and due to the line equivalent width, most of the observed signal should be above zero throughout the 41.5 MHz frequency range of the original voxel.

\end{itemize}

In addition to the promising prospects for follow up by external surveys, the statistically significant number of CO voxels individually identified can be used to constrain the overall molecular gas content of old, gas rich galaxy clusters.
 
\subsection{CO equivalent width : COMAP + HETDEX}
\label{subsec:Analysis.3}
\begin{figure}
    \includegraphics[width=\columnwidth]{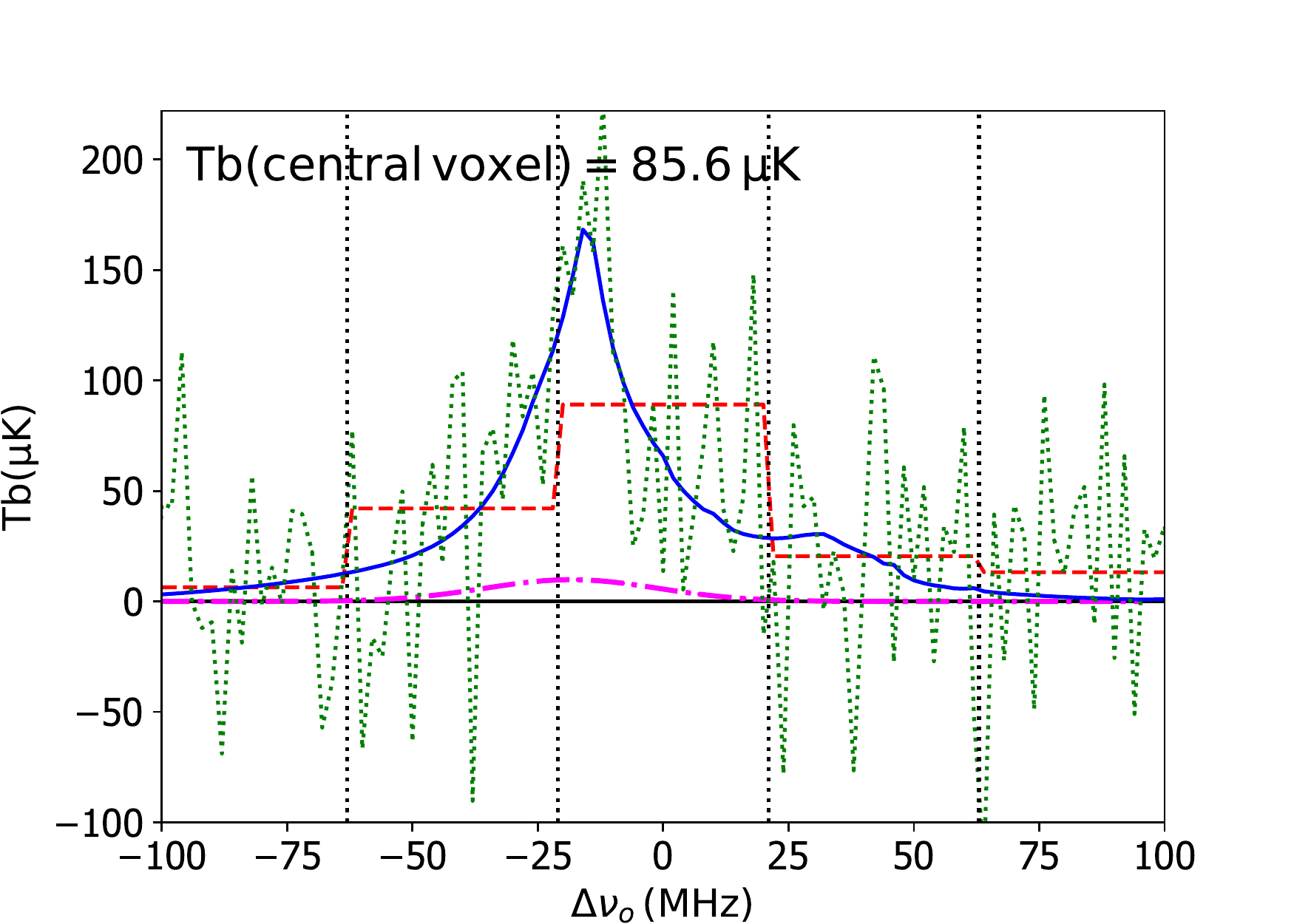}
\caption{CO brightness temperature around a COMAP bright voxel, with a particularly high signal, obtained using a 2 MHz spectral resolution. The green dotted line and red dashed line show the observed signal (including noise) with respective the high and low frequency resolution, the blue solid line shows only the signal. Black vertical lines indicate the limits of a voxel with the COMAP.H spectral resolution. The central voxel was identified following the description from Section~\ref{subsec:Analysis.2}. The black solid line marks Tb = 0. The magenta dashed dotted line corresponds to the signal from the brightest galaxy in the voxel ($L^{\prime}_{\rm CO}=4.6 \times \, 10^{11}\, ({\rm K\, km\,  s^{-1}\, pc^2})$). This bright galaxy only contributes with a few percent of the total observed signal. }    
\label{Fig:7}
\end{figure}

As discussed in Section~\ref{subsec:EW} the signal from bright CO emitters will have a large EW and so it will often be observed over several voxels in the spectral direction; see Figure~\ref{Fig:7} for a visualization of a bright CO emitter with a wide EW. This plot shows that the signal from a galaxy can be extended throughout several voxels but that the overall observed signal and shape is a convolution from the signal of a series of galaxies (extreme bright galaxies are often found in clusters).

The signal in bright CO voxels, together with the signal from neighboring voxels in the spectral direction, re-scaled to a higher frequency resolution can be used to probe the equivalent width of CO emitters.

We adopted a new frequency binning of $2\, {\rm MHz}$ which translates to an average noise level of $53.3\, \mu {\rm K}$ and analyzed the bright voxel emission individually and using stacking. Each selected bright structure was analyzed over 104 ($2\, {\rm MHz}$) voxels (corresponding to five $41.5\, {\rm MHz}$ voxels).

If we assume that a galaxy CO line width is driven by thermal broadening, then its maximum EW  can be theoretically estimated from the maximum halo mass as a function of redshift. COMAP bright voxels will be inspected to look for particularly large EWs which would be a possible indicator of turbulent gas.

\begin{figure}
    \includegraphics[width=\columnwidth]{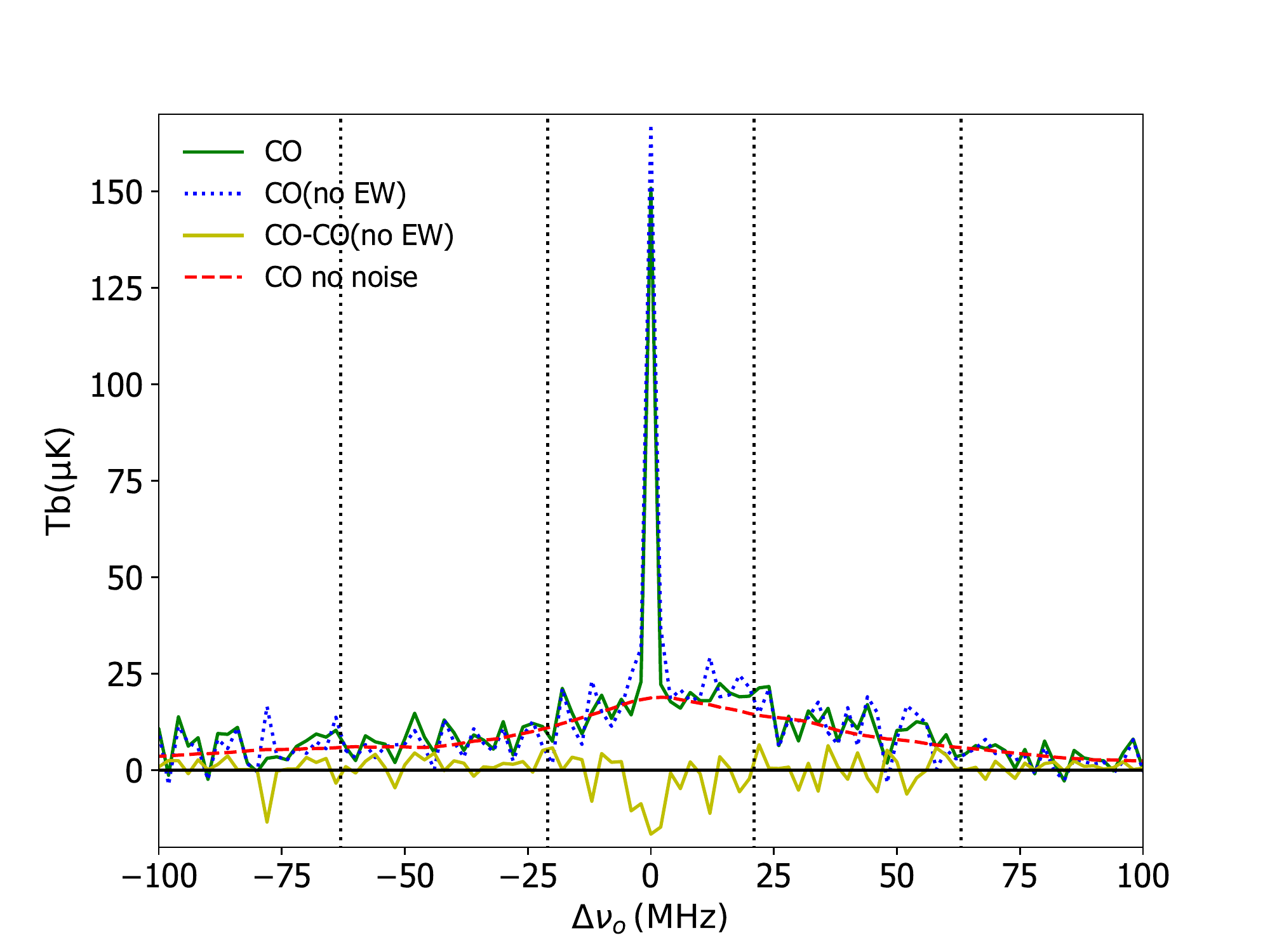}
\caption{CO brightness temperature of a stack of COMAP bright voxels obtained using a 2 MHz spectral resolution. The stack is comprised of all the bright voxels identified following the description from Section~\ref{subsec:Analysis.2} over one COMAP field, except for those which can be visually identified as noise peaks; the position $\Delta \nu_0$ corresponds to the position of the brightest peak (signal + noise) in a bright voxel. The green solid line and blue dotted line shows the observed signal (including noise) assuming respectively the fiducial broad line width and no line width corrections. The yellow solid line shows the change in observed signal when the galaxies equivalent width is accounted for.}    
\label{Fig:8}
\end{figure}

We tested the voxel EW in our simulations and found that:

\begin{itemize}
\item The number of bright voxels detected following the procedure described in \ref{subsec:Analysis.2} would increase by around 40\% if the CO emitters had narrow equivalent widths (i.e.\ well below the survey resolution).
\item In the cases where the bright voxel signal is highly dominated by one or two neighboring galaxies, the S/N of the detection increases in neighboring voxels (in the spectral direction).

\item When we compare the signal in voxels adjacent to bright voxels (identified as in \ref{subsec:Analysis.2}) in the spectral direction with those in the direction perpendicular to the line of sight, we find that the signal is on average 2.5 to 3 times higher in the spectral direction. Since COMAP spatial resolution in the spectral direction is much higher this result indicates that we are detecting extended structures or/and structures with large EWs. This is an expected result given that CO emission traces the large scale structure and so we will observe emission from extended clustered regions.

\item For the stacking case (of all the bright voxels) we find that when the galaxies EW is taken into account, the signal from the brightest voxels is decreased compared to adjacent voxels in the spectral direction (see Figure~\ref{Fig:8}). This figure also shows that the signal in voxels adjacent to the identified bright voxel is due to a mixture of EW effects, the number of CO emitters in the region and the size of the emitting regions (in the spectral direction).
If all of the emitting galaxies were located in the center of the voxel and if there was no significant emission from nearby voxels then the excess signal in adjacent voxels would make it possible to probe the average EW of the CO emission from the galaxies.

\item For extreme bright sources we will be able to probe the EW of an individual galaxy, such as in the case illustrated in Figure~\ref{Fig:8}. 
\item For most galaxies and galaxy clusters, we can only provide the probability that the bulk emission in these voxels has a combination of size plus CO EW comparable to or above the voxel spectral width. As shown in Figure~\ref{Fig:8} the stack of the spectra of several bright sources is smoothed due to the broadness of the CO lines. The stack shown in  Figure~\ref{Fig:8} contains the bright source candidates after removing sources which by visual inspection appear to be noise dominated. Moreover, we adjust the peak of the signals so to maximize the visible effect of the EW in increasing the overall brightness of a series of voxels. 
\item We note that Figure~\ref{Fig:8} illustrates well the smoothing effect on CO peaks that will affect the CO power spectra, as discussed in detail in \cite{2021arXiv210411171C}.

\end{itemize}

\subsection{Upper limit on Total CO emission at z=6: COMAP + HETDEX}
\label{subsec:Analysis.4}

By using the HETDEX-detected LAEs as tracers of large scale structure at $z\sim 3$ we can evaluate the signal in COMAP maps and try to isolate the  ${\rm CO}(1-0)$ from the  ${\rm CO}(2-1)$ signal. This classification assumes that ${\rm CO}(1-0)$ emission is correlated with LAE rich regions in the same volume (signaling an overdense region such as a galaxy cluster). On the other hand, ${\rm CO}(2-1)$ bright voxels emitted from EoR galaxies and sky noise are uncorrelated with HETDEX data.

In summary, we used the positions of the LAEs to mask most of the low redshift emission and derived an upper limit on the high redshift signal by stacking the remaining voxels.

\begin{figure*}
\begin{center}
\includegraphics[width=18.4cm]{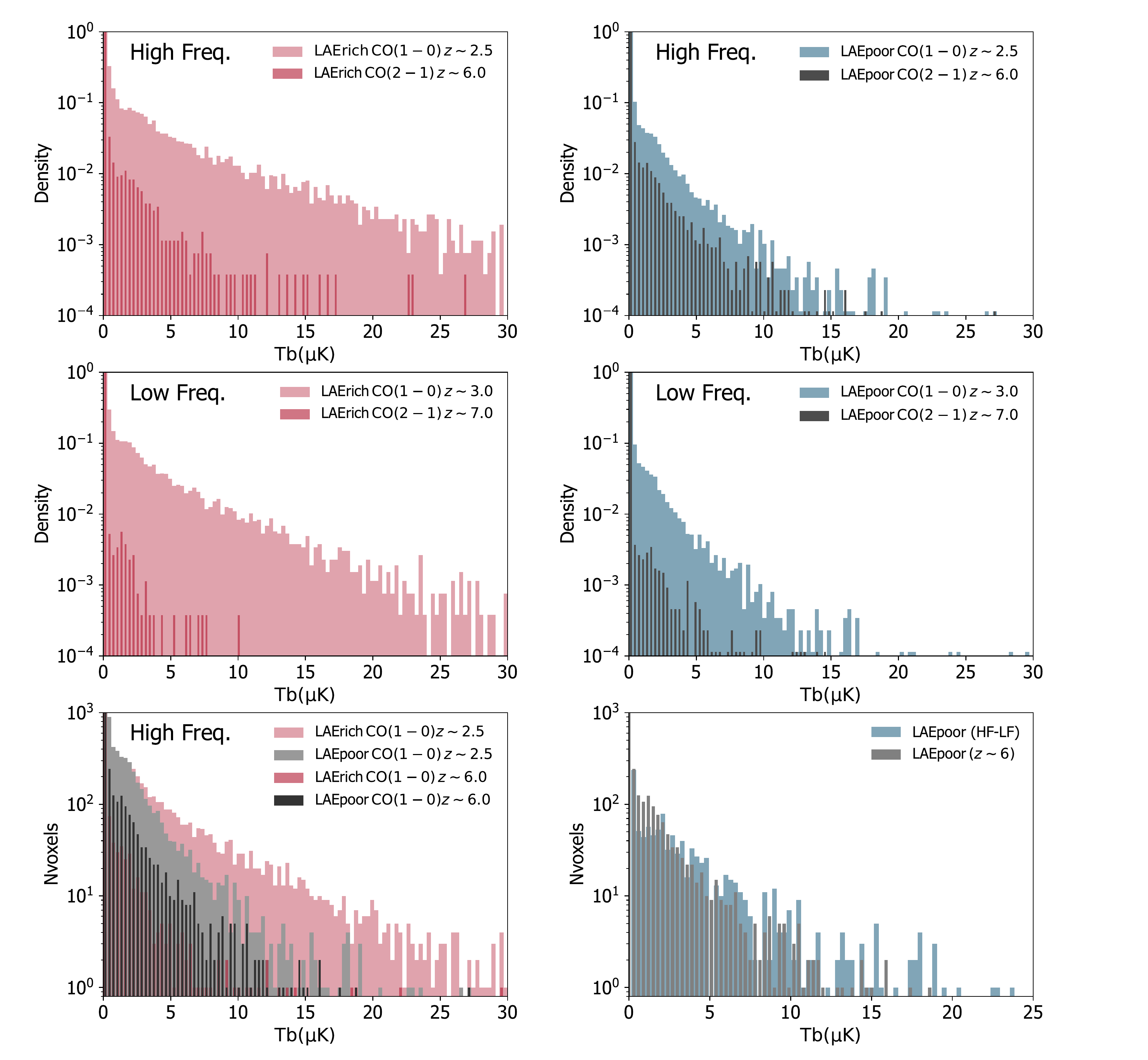}
\end{center}
\caption{VID of ${\rm CO(1-0)}$ and ${\rm CO(2-1)}$ from the volume overlapping with LAE rich regions (${\rm LAErich}$) and from the remaining voxels of the COMAP map (${\rm LAEpoor}$), separated in LF and HF data-sets. Top left panel shows the VID in the (${\rm HF, LAErich}$) volume where the low redshift ${\rm CO(1-0)}$ line greatly dominates the overall signal; this volume contains $\sim85\%$ of the total ${\rm CO(1-0)}$ average $T_{\rm CO(1-0)}({\rm LAErich})\sim 2 T_{\rm CO(1-0)}({\rm LAErich}+{\rm LAEpoor})$. Top right panel shows the VID in the (${\rm HF, LAEpoor}$) volume where both the low redshift ${\rm CO(1-0)}$ and the high redshift ${\rm CO(2-1)}$ make a meaningful contribution to the total signal. Bottom left panel contains the four HF VID's from the two lines and the two volumes. This plot illustrates how much the signal from the ${\rm CO(1 - 0)}$ line changes between the volumes ${\rm LAErich}$ and ${\rm LAEpoor}$ whereas the ${\rm CO(2-1)}$ remains unchanged. In the simulations used in this plot ${\rm LAErich}$ and ${\rm LAEpoor}$ contain respectively $45\%$ and $55\%$ of the total map.
The evolution of the PDFs from the top to the middle panels illustrates the small change in the ${\rm CO(1-0)}$ signal compared to that of the ${\rm CO(2-1)}$ signal between the HF and LF maps maps. Bottom, right panel compares the difference in the VID of the HF and LF maps with the VID of ${\rm CO(2-1)}$ at $z\sim6$.}  
\label{Fig:9}
\end{figure*}

We started by dividing the HF map in two volumes. The first, ${\rm LAErich}$ contains all the HETDEX bright voxels and voxels adjacent to HETDEX bright voxels in the spectral direction. The second, ${\rm LAEpoor}$ contains the remaining voxels. Note that we are only using the high frequency maps $({\rm HF} ={\rm 30\, -\, 34\, GHz})$ since the ${\rm CO}(2-1)$ signal is expected to decrease drastically from $z\sim6$ to $z\sim7$. Figure~\ref{Fig:9} shows the expected signals from each line in the different volumes.

We found that:
\begin{itemize}
\item The volume ${\rm LAErich}$ contains $\sim 45\%$ of COMAP voxels and the average ${\rm CO}(1-0)$ in ${\rm LAEpoor}$ is lower by a factor of $\sim 4$ compared to the total map $\left(T_{\rm CO(1-0)}({\rm LAEpoor})=0.25\, T_{\rm CO(1-0)}\right)$.
\item The signal in volume ${\rm LAEpoor}$ yields an upper limit on the high-$z$ ${\rm CO(2-1)}$ signal.
\item A rough estimation of the significance of the ${\rm CO(2-1)}$ upper limit that can be derived from this method can be made assuming that the real signals amplitude is $T_{\rm CO(1-0)}=\, 0.8\, \mu {\rm K} $ and $T_{\rm CO(2-1)}=\, 0.1\, \mu {\rm K} $. Which implies that we would be able to derive an upper limit on the high-$z$ signal of $T_{\rm CO(2-1)}<0.3\, \mu {\rm K}$. Note that the significance of the upper limit will increase with the ratio $T_{\rm CO(2-1)}(z~\sim~6)/T_{\rm CO(1-0)}(z~\sim~2.5)$ and this study predicts a particularly steep decrease of the CO emission towards high-$z$.
\item We estimate a signal significance of $\sim 3.7\sigma$ over one field and $\sim 6.5\sigma$ over three fields.
\end{itemize}

Note that these predictions are model dependent and so the constraints in the ${\rm CO(2-1)}$ emission for our different simulations vary between an upper limit of (2.5 to 4) times the real $T_{\rm CO(2-1)}$. In any case, this signal is currently uncertain by one to two orders of magnitude and so this would be a relevant upper limit. A large $T_{\rm CO(2-1)}(z\sim6)$ (detectable using the described procedure) would imply a combination of large molecular gas reservoirs in EoR galaxies and evolution of galaxies average interstellar medium properties leading to molecular gas temperatures well above the CMB temperature; this scenario is favored by observations by the COLDz survey at $z\sim 5.8$. On the other hand, a non detection would imply a low CO signal, or a particularly anti-correlation between galaxy clusters hosting LAEs and CO bright galaxies.

\subsection{Estimate of the Total CO emission at z=6: COMAP + HETDEX}
\label{subsec:Analysis.5}

Given that the ${\rm CO(1-0)}$ signal is much stronger than the ${\rm CO}(2-1)$ signal the upper limit derived with the method described in Section~\ref{subsec:Analysis.4} cannot exclude the possibility of an extremely low ${\rm CO}(2-1)$ emission. Therefore, we now describe how to get a more meaningful  constraint on the ${\rm CO}(2-1)$ signal at $z\sim6$.

The intensity of the target CO lines is expected to decrease towards high-redshift (over $2<z<8$) due to the decrease in galaxy number density, lower dust/metallicity and higher CMB temperature (which decreases the contrast between $T_{\rm CO}$ and $T_{\rm CMB}$). This decrease is expected to be especially significant at $z>7$.
Therefore, the percentage evolution of the ${\rm CO}(1-0)$ signal probed by COMAP over $z = 2.4$--3.4 should be much lower than that of the ${\rm CO}(2-1)$ signal over $z = 5.8$--7.9. This signal decay from low to high redshift is expected with a high level of confidence independently of the signals amplitude.

Based on this expectation, we derive our constraint by dividing COMAP data in two, based on frequency, with the 'low-frequency' subset consisting of frequencies in the range 26--30\,GHz and the 'high-frequency' subset consisting of frequencies in the range 30--34\,GHz. The HF data-set therefore contains emission from ${\rm CO}(1-0)$  ($2.39 <z < 2.85$) and ${\rm CO}(2-1)$ at ($5.78 <z < 6.84$) while the LF data-set contains ${\rm CO}(1-0)$  at ($2.85 <z < 3.43$) and ${\rm CO}(2-1)$ at ($6.84 <z < 7.87$). Both the HF and LF data-sets are themselves divided in two volumes according to their overlap with HETDEX bright voxels following the prescription outlined in  Section~\ref{subsec:Analysis.4}.

The CO data is therefore divided into four volumes: (${\rm HF,  LAErich}$), (${\rm HF, LAEpoor}$), (${\rm LF, LAErich}$), (${\rm LF, LAEpoor}$) (see Figure~\ref{Fig:9}). We assume the size of (${\rm HF, LAErich}$) and (${\rm HF, LAEpoor}$) as $45\%$ of the total map (an estimate obtained from our simulations).

By analyzing COMAP mock HF and LF data-set we found that:
\begin{itemize}
\item We expect a decrease of $5-20\%$ in the ${\rm CO}(1-0)$ signal and $70-80\%$ in the ${\rm CO}(2-1)$ signal from the HF to the LF map. As default we assume a reduction of $10\%$ and $75\%$ respectively for ${\rm CO}(1-0)$ and ${\rm CO}(2-1)$.
\item With the observational data, the percentage evolution of the ${\rm CO(1-0)}$ line can be estimated as (${\rm HF, LAErich}$) - (${\rm LF,  LAErich}$) and the percentage evolution of the ${\rm CO(2-1)}$ line can be estimated as (${\rm HF, LAEpoor}) - ({\rm LF, LAEpoor}$).
\item A rough estimation of the contribution of each line to each data-set can be made assuming that the initial signals in the HF map are $T_{\rm CO(1-0)}({\rm HF})=\, 0.8\, \mu {\rm K} $ and $T_{\rm CO(2-1)}({\rm HF})=\, 0.1\, \mu {\rm K} $ and $\left(T_{\rm CO(1-0)}({\rm HF,{\rm LAEpoor}})=0.25\, T_{\rm CO(1-0)}({\rm HF})\right)$. It follows that $T_{\rm CO(1-0)}({\rm HF},{\rm LAEpoor})=\, 0.20\, \mu {\rm K}$, $T_{\rm CO(1-0)}({\rm LF},{\rm LAEpoor})=\, 0.18\, \mu {\rm K}$, $T_{\rm CO(2-1)}({\rm HF},{\rm LAEpoor})=\, 0.10\, \mu {\rm K}$ and $T_{\rm CO(2-1)}({\rm LF},{\rm LAEpoor})=\, 0.02\, \mu {\rm K}$. From what follows that our estimate on the ${\rm CO(2-1)}({\rm HF})$ line is $\left({\rm HF,\, LAEpoor}\right) - \left({\rm LF,\, LAEpoor}\right)\, =\, 0.10\, \mu {\rm K}$.  The precision of this estimate increases with the ratio between the high-$z$ and the low-$z$ CO signals.
\item Note that the ${\rm CO}(1-0)$ and ${\rm CO}(2-1)$ lines changed respectively by $0.02\, \mu {\rm K}$ and $0.08\, \mu {\rm K}$ from the (${\rm HF, LAEpoor}$) to the (${\rm LF, LAEpoor}$,) map. Given that the rate of redshift evolution of the CO emission at $z\sim2$-- 3 should be small or even close to zero and that at $z\gtrsim6$ the CO signal will decay drastically this is a very robust method to probe the $z\sim6$ CO signal. 
\item
Based on the previous assumptions we find that the observed difference in the intensity between COMAP (${\rm HF, LAEpoor}$) and (${\rm LF, LAEpoor}$) maps is therefore $\Delta T_{\rm CO}=(0.2+0.1)-(0.18 + 0.02)\,=\, 0.1\,  \mu {\rm K}$. Note that the ${\rm CO}(1-0)$ and ${\rm CO}(2-1)$ lines decreased respectively by 0.02 $\mu {\rm K}$ and 0.08 $\mu {\rm K}$.
\end{itemize}

The expected evolution in the two lines between the LF and HF maps is illustrated in Figure~\ref{Fig:9}.

\subsection{Voxel intensity distribution (VID) with COMAP + HETDEX}
\label{subsec:Analysis.6}

The distribution of CO intensities over COMAP voxels can be estimated by assuming that it is responsible for the deviation from the Gaussian-like distribution expected for the noise peaks. The voxel intensity distribution (VID) in COMAP maps was discussed in \cite{2019ApJ...871...75I} as a statistic that can be used to probe the CO luminosity function. Given the large volume in a COMAP voxel, VID luminosity functions differ from galaxy luminosity functions because several galaxies contribute to the emission measured in each voxel. Nevertheless, we can extract similar information from both statistical measurements.

The use of COMAP VID data to probe emission from possible Dark Matter (DM) decays over the COMAP frequency range and detect or exclude some of the free parameter space for weakly interactive DM candidates such as the Axion was discussed in \cite{2020arXiv201200771B}.

Here we explore the possible application of the VID separately to COMAP voxels in regions bright or faint in LAEs (respectively referred to as ${\rm LAErich}$ and ${\rm LAEpoor}$) as an additional way to probe the CO emission in these maps or to detect any other signals.
COMAP data are highly dominated by ${\rm CO}(1-0)$ line emission from $z\sim3$. By partitioning COMAP data into HETDEX bright and HETDEX faint volumes we can compare how well CO and LAEs are spatially correlated just by comparing the CO luminosity density in each data-set.

\begin{itemize}
\item In the simulated data we find that using the VID we can probe which range in CO brightness is being identified using the HETDEX bright voxels.
\item By changing the selection to identify HETDEX bright voxels and later applying the VID to the derived data-sets we can optimize the selection to maximize the probability of identifying a particular type of galaxy.
\item For ${\rm LAErich}$ data (LF and HF), the VID will be highly dominated by ${\rm CO}(1-0)$ emission. However, for ${\rm LAEpoor}$ data, the VID for HF will have a large contribution from the EoR ${\rm CO}(2-1)$ signal that is expected to be much reduced for LF (see Figure~\ref{Fig:9}). A comparison of the ${\rm LAEpoor}$ HF and LF VID shapes will reveal this difference.
\item The EoR CO emission will be mainly distributed over relatively faint CO voxels.
If this signal is particularly bright then the VID in HF ${\rm LAEpoor}$ voxels can be used to determine how much of the signal originates from bright and clustered versus faint sources.
\item A large contribution to the EoR CO emission from relatively faint sources would be an indicator that the future ngVLA will only detect a fraction of the overall EoR CO emission.
On the other hand, an EoR signal dominated by bright sources would imply that there is little CO emission in most EoR galaxies due to a combination of low metallicity and low gas temperature.
\item The LF ${\rm LAErich}$ maps are ideal to probe the ${\rm CO}(1-0)$ VID, given that the signal in this volume is about twice as bright as the signal from all the HF maps which makes it easier to distinguish signal from noise. Moreover in this frequency range the contamination by the EoR ${\rm CO}(2-1)$ line emission is minimal.
\item As illustrated in the bottom right panel of Figure \ref{Fig:9} the difference between the VID of the HF and LF ${\rm LAEpoor}$ maps will likely provide a rough estimate estimate of the VID of ${\rm CO}(2-1)$ at $z\sim6$. 
The data separation in the two volumes was based only on the external HETDEX LAEs data and so the observational noise (not shown in the VIDs) in each volume will remain as Gaussian as the initial map. Assuming that the noise from the observational map is Gaussian enough each of the two volumes has enough voxels for the VID to be detected as an excess above the noise. If the overall data indicate that this ``difference'' VID is truly dominated by the EoR line then this result will have important implications for future EoR CO missions and for the physical properties of the ISM of EoR galaxies. Even in a scenario where this difference VID is dominated by the low redshift emission it would imply a low EoR signal; such a result would on its own have significant implications. The most pessimistic case would be a low significance measurement of the difference PDF due to noisy data which would still be useful as a sanity check for the overall maps analysis.
\item An upper limit on a signal from dark matter decays, such as that proposed by \citep{2020arXiv201200771B}, can be made using the ${\rm LAEpoor}$ maps at higher sensitivity than if we had used the total map.
\end{itemize}

From our simulations we conclude that the VID over HETDEX bright/faint voxels can be used to probe the CO(1-0) brightness distribution if some correction for the EW is applied and if the noise in the brightest COMAP voxels is close to Gaussian. This exercise can at least provide a probe of the average brightness of CO emitters traced or missed by HETDEX LAEs and can be used to optimize data selection.
For the case of the high-$z$ ${\rm CO}(2-1)$ emission, the overall brightness of the voxels containing the bulk emission is quite low and therefore this signal will be sensitive to the noise Gaussianity in relatively faint voxels ($\lesssim 7 \mu {\rm K}$).

\section{Bright voxel spectra analysis and observational effects}
\label{sec:Obs_eff}

By applying the analysis techniques outlined in Section~\ref{sec:Analysis} to the COMAP data we expect to obtain a catalog with spectra of hundreds of bright CO structures (with a 2 MHz resolution).

Through careful individual examination, noisy spectra can be excluded from the catalog and some tentative spectral sizes can be measured for different sources. The spectra of these sources are in some cases a blending of the emission from several clustered regions, but they can often be identified by visual data inspection (if the sources are separated in the line of sight direction). 
The frequency extent of the bright structures will depend on 1) the galaxies' EW, 2) their size and 3) the number of galaxy structures in the sample. In addition, the conversion of angular positions and frequencies into distances will be affected by ``observational effects'' such as 4) peculiar velocities and 5) the Alcock-Paczynski effect. Finally, we need to make assumptions about non-Gaussianity to derive a possible two-point correlation function and test the Alcock-Paczynski effect.

We now briefly outline how we plan to account for these effects in our analysis:

\begin{enumerate}
\item In Section~\ref{subsec:Analysis.3} we discussed the effect of the EW in observations. The EW of a source will depend on the size of the gravitationally bound structure containing it and it will increase in the presence of non-thermal gas turbulence. CO lines are associated with broad EWs which should scale with the galaxy luminosity. For galaxy clusters the observed EW is a convolution of the cluster size and the EW of each of its components. We find that for the case of COMAP.H bright voxels, each voxel contains one to 3/4 galaxy clusters which translates to an overall broadness that can range between $\sim$25 to $\sim$200 MHz (for Tb $\gtrsim 3\, \mu{\rm K}$).

\item If we ignore peculiar velocities and uncertainty effects due to assumptions about the cosmology then the spectral size of the source would translate into its physical size (for lines with a narrow EW). In the case of COMAP, sources sizes, galaxy clusters numbers and EWs will be estimated together. 

\item Peculiar velocities will cause the elongation of the observed structures in the spectral direction. COMAP's spatial resolution limits our ability to measure the transverse size of these structures and so we cannot easily measure or correct for this effect. However, we will add the possible effect of peculiar velocities to the error bars of the structures' sizes.

\item The Alcock-Paczynski effect can be used to test the accuracy of the conversion from angles and frequencies to distances. By forcing isotropy in the correlation function from radial (along the line of sight) and transverse distances between objects we can derive constraints on the product ${\rm H(z)D_A(z)}$. COMAP data has enough frequency resolution $(d_{ \parallel\perp} \sim 0.27\, {\rm cMpc}\, (z \sim 3)$) to measure radial distances but not enough angular resolution to measure precise transverse distances $(d_{\perp} \sim 8.5\, {\rm cMpc}\, (z \sim 3))$. However, if we assume that the position of HETDEX LAEs (1.5 arcsec resolution) provides a good estimate for the structure position on the plane perpendicular to the line of sight, then the Alcock-Paczynski effect can be tested with the data.  The latter effect, although challenging to apply to the COMAP plus HETDEX data-sets, would be interesting to detect. We will evaluate the feasibility of such a test from the observational data.
We note that the Alcock-Paczynski effect will not affect our study as long as we assume a consistent cosmology for the HETDEX and COMAP data. 

\end{enumerate}

We plan to account for all of the previously described effects in COMAP data during the analysis of the observational data. The first three, CO line width, cluster size and source blending, were accounted-for in this study. We expect the effect from peculiar velocities to be minimal for data analysis using 41.5 MHz resolution but will nevertheless account for it as a possible error source. The Alcock-Paczynski effect will have a very minimal impact on the voxel-based COMAP analysis. 

\section{Summary and Conclusions} \label{section }

In this study we used a series of self consistent models for CO and Lyman alpha emission to produce mock observations for COMAP CO maps and to derive a mock LAE catalog for the HETDEX surveys. The different models span the likely uncertainty in the CO signal and account for the lines' dependence on dust emission. 

We used the simulated data to explore a series of different ways in which HETDEX LAE catalogs can be used both to increase the sensitivity of a CO detection by COMAP and to interpret the CO data. HETDEX data will be particularly useful to identify signal-dominated voxels (in the brightest COMAP voxels) and potentially to constrain the ${\rm CO}(2-1)$ signal from EoR structures.

Our study indicates that a series of major science goals will be enabled by synergies between the COMAP/HETDEX missions. We expect to:\\
i) Increase the sensitivity of an initial ${\rm CO}(1-0)$ line detection by a factor of two to three.\\
ii) The high signal-to-noise ratio expected for a CO detection will make it possible to divide COMAP data into several redshift bins and thus probe the redshift evolution of the signal in the $2<z<8$ range.\\
iii) Individually identify hundreds of COMAP voxels with bright ${\rm CO}(1-0)$ emission from gas-rich galaxy clusters at $ z\sim3$. Galaxy clusters at $ z>2$ are typically detected using optical/UV lines which introduces a bias towards low dust (low extinction) systems. COMAP data will provide a complementary view of these clusters, free of this particular bias.\\
iv) Probe the equivalent width from the brightest clusters in hundreds of COMAP voxels, providing a statistical sample to explore for evidence of gas turbulence. \\
v) Shed light on how the sources of CO and LAEs are spatially correlated. CO emission is associated with old dusty galaxies while LAEs are more common in young dust-free galaxies. The VID from LAE-rich versus poor regions will be significantly different since CO-emitting galaxies and LAE are correlated at the cluster scale (hence, they are usually correlated at the scale of a COMAP.H voxel). \\
vi) Strongly improve upon the current limits on the ${\rm CO}(2-1)$ emission at $z\sim6$. HETDEX LAE data can be used to mask most of the ${\rm CO}(1-0)$ signal from COMAP data which will make it possible to put a $\sim6\sigma$ upper limit on the residual ${\rm CO}(2-1)$ signal for three fields ($\sim3\sigma$ for one field). Note that the combination of these two data-sets can provide an upper limit/estimation of the ${\rm CO}(2-1)$ emission down to a level of $\sim 0.05\,  {\rm \mu K}$. Such a low signal would have strong implications for future efforts to detect ${\rm CO}(1-0)$ emission from EOR structures, for example using a future COMAP-{\em EoR} survey (targeting the Ku band 13--17 GHz, see Breysse et al. in prep.) or the ngVLA. In addition a particularly low signal would constrain the dust content, gas temperature and overall properties of the interstellar medium of EOR galaxies. \\
vii) Probe the distribution of bright temperatures in CO voxels that could be compared with ${\rm CO}(1-0)$ luminosity functions at $z\sim2.4$ by the COLDz galaxy survey to test the compatibility of the two.\\
viii) Detect or provide independent/competitive upper limit to signals from dark matter decay in the COMAP frequency range.

In summary, the combination of COMAP and HETDEX data has the potential to bring a better understanding to the drivers of the SFRD at $z\sim2-3$, to the CO content of LAEs at high-$z$ and to properties of the first galaxies during the EOR.

The future for LIM experiments lies in the combination of information from different LIM observations targeting multiple lines as well as from other surveys. We have presented a series of methods that can be used to interpret COMAP CO LIM data by making use of information from an external LAE catalog. Although designed specifically this purpose, these methods have wide application to other combinations of LIM and external surveys in general and will become increasingly important as information from multiple overlapping LIM experiments becomes available. 

\begin{acknowledgements}
BB acknowledges support from the Research Council of Norway through NFR Young Research Talents Grant 276043. KC acknowledges funding from NSF Awards 1518282 and 1910999.
DTC is supported by a CITA/Dunlap Institute postdoctoral fellowship. The Dunlap Institute is funded through an endowment established by the David Dunlap family and the University of Toronto.
HP acknowledges support from the Swiss National Science Foundation through Ambizione Grant PZ00P2\_179934. LCK was supported by the European Union’s Horizon 2020 research and innovation programme under the Marie 
Skłodowska-Curie grant agreement No. 885990. JK is supported by a Robert A. Millikan Fellowship from Caltech.
COMAP is supported by funding from NSF Award 1910999.
\end{acknowledgements}
\bibliography{bibliography}{}
\bibliographystyle{aa} 

\end{document}